\documentclass{article}

\usepackage{arxiv}

\usepackage[utf8]{inputenc} 
\usepackage[T1]{fontenc}    
\usepackage{hyperref}       
\usepackage{url}            
\usepackage{booktabs}       
\usepackage{amsfonts}       
\usepackage{nicefrac}       
\usepackage{microtype}      
\usepackage{lipsum}		
\usepackage{graphicx}
\usepackage{natbib}
\usepackage{doi}
\usepackage{caption}
\usepackage{subcaption}
\usepackage{algorithm}
\usepackage{algorithmic}
\usepackage{amsmath}

\title{Recursive Camera Painting: A Method for Real-Time Painterly Renderings of 3D Scenes}

\date{} 					

\author{ 
 \href{https://orcid.org/0000-0003-3618-4166}{\includegraphics[scale=0.06]{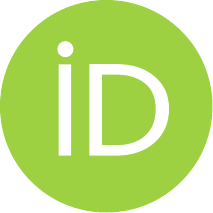}\hspace{1mm}Ergun Akleman}\thanks{Joint with Computer Science and Engineering Department.} \\
	Visual Computing \& Computational Media,\\ Texas A\&M University, College Station, TX, 77831\\
	\texttt{ergun@tamu.edu} \\
  \And
  \href{https://orcid.org/0000-0002-5092-7061}{\includegraphics[scale=0.06]{orcid.pdf}\hspace{1mm}Cassie Mullins}\\
    Department of Visualization\\ 
    Texas A\&M University, College Station, TX, 77831\\
    \texttt{cmullins7@tamu.edu} \\
   \And
	 Christopher Morrison\\
Department of Visualization\\ 
Texas A\&M University, College Station, TX, 77831\\
	\texttt{cmorr@tamu.edu} \\
    \And
	 David Oh\\
Department of Visualization\\ 
Texas A\&M University, College Station, TX, 77831\\
	\texttt{somyungoh@tamu.ed} \\
}

\renewcommand{\shorttitle}{Recursive Camera Painting}

\hypersetup{
pdftitle={A template for the arxiv style},
pdfsubject={q-bio.NC, q-bio.QM},
pdfauthor={David S.~Hippocampus, Elias D.~Striatum},
pdfkeywords={First keyword, Second keyword, More},
}

\begin{document}
\maketitle

\begin{abstract}
In this work, we present the recursive camera-painting
approach to obtain painterly smudging in real-time rendering applications. We have implemented recursive camera painting as both a GPU-based ray-tracing and in a Virtual Reality game environment. Using this approach, we can obtain dynamic 3D Paintings in real-time. In a camera painting, each pixel has a separate associated camera whose parameters are computed from a corresponding image of the same size. In recursive camera painting, we use the rendered images to compute new camera parameters. When we apply this process a few times, it creates painterly images that can be viewed as real-time 3D dynamic paintings. These visual results are not surprising since multi-view techniques help to obtain painterly effects. 
\end{abstract}

\section{Introduction}

 \begin{figure}[htbp!]
    \centering
\includegraphics[width=0.49\linewidth]{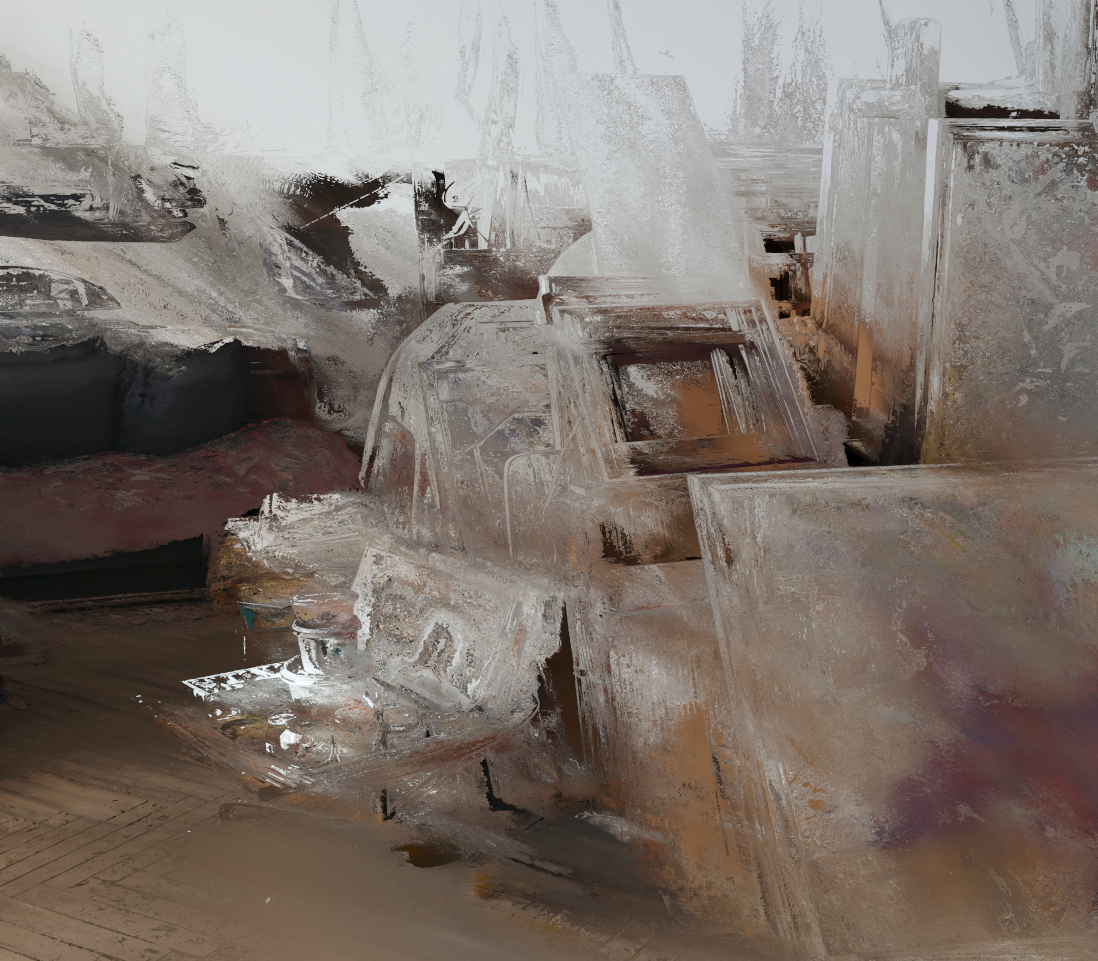}
\hfill
\includegraphics[width=0.49\linewidth]{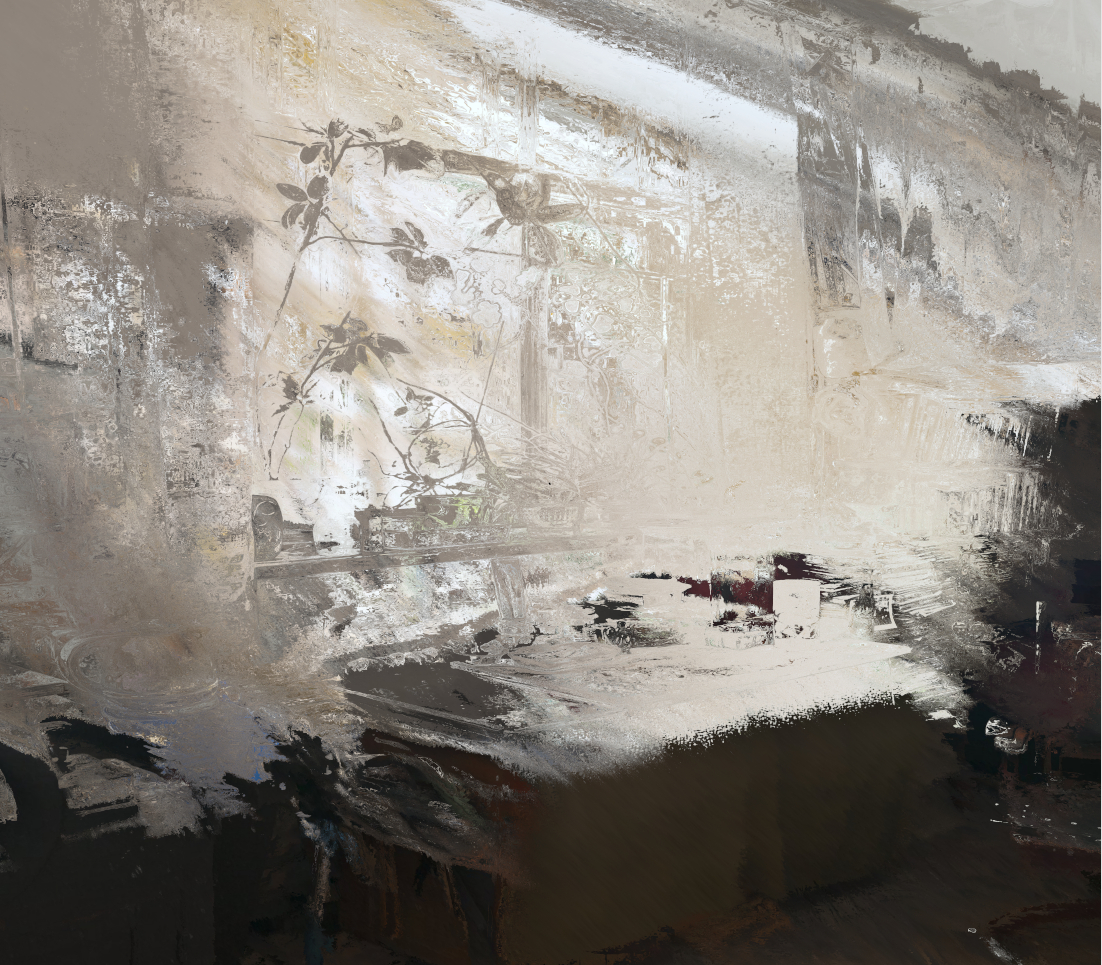}
\hfill
\vspace{0.05in}
\includegraphics[width=0.49\linewidth]{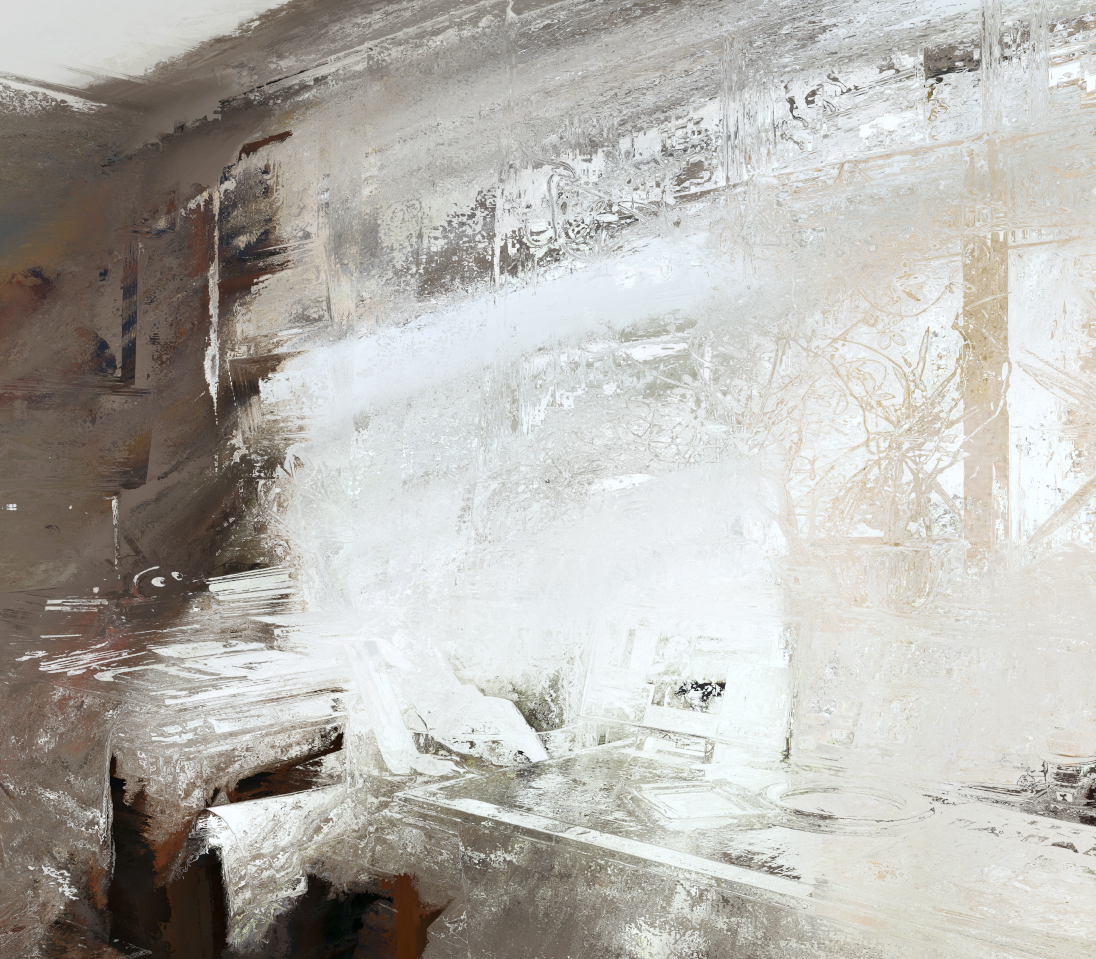}
\hfill
\includegraphics[width=0.49\linewidth]{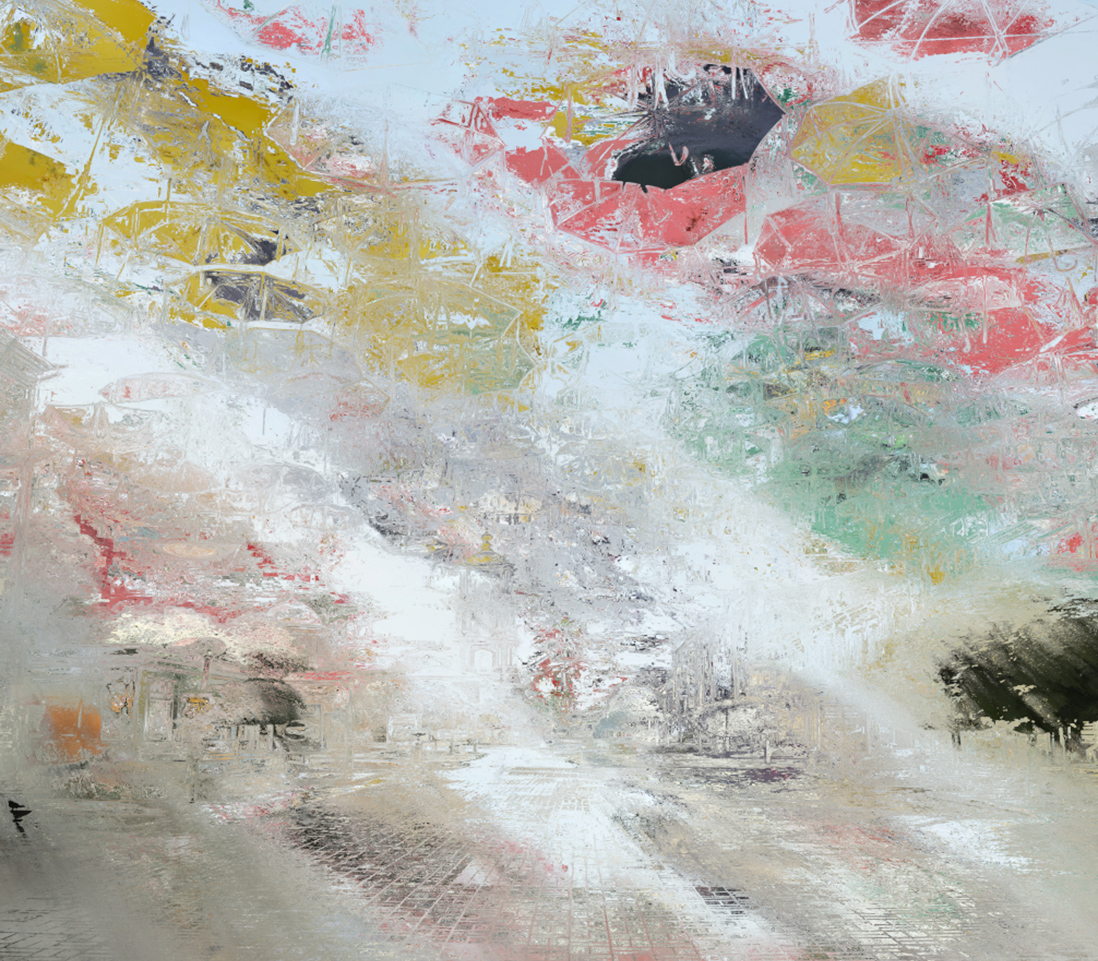}
\caption{\it Examples of images created in real-time with our approach implemented as a GPU-based ray tracing system. }
\label{fig_teaser}
\end{figure}

During the end of the nineteenth century, traditional perspective conventions were broken down in art. From Cezanne, artists quickly discovered that the multi-view perspective is an important tool for obtaining visually appealing images. Not only traditional cubists, but also a wide variety of artists such as Pablo Picasso, Willem de Kooning, and David Hockney from a variety of art schools employed the multi-view perspective for the creation of expressive depictions \cite{hess2004willem,hockney1995david}.

\begin{figure*}[thbp!]
    \centering
    \begin{subfigure}[t]{0.46\linewidth}
\includegraphics[width=\linewidth]{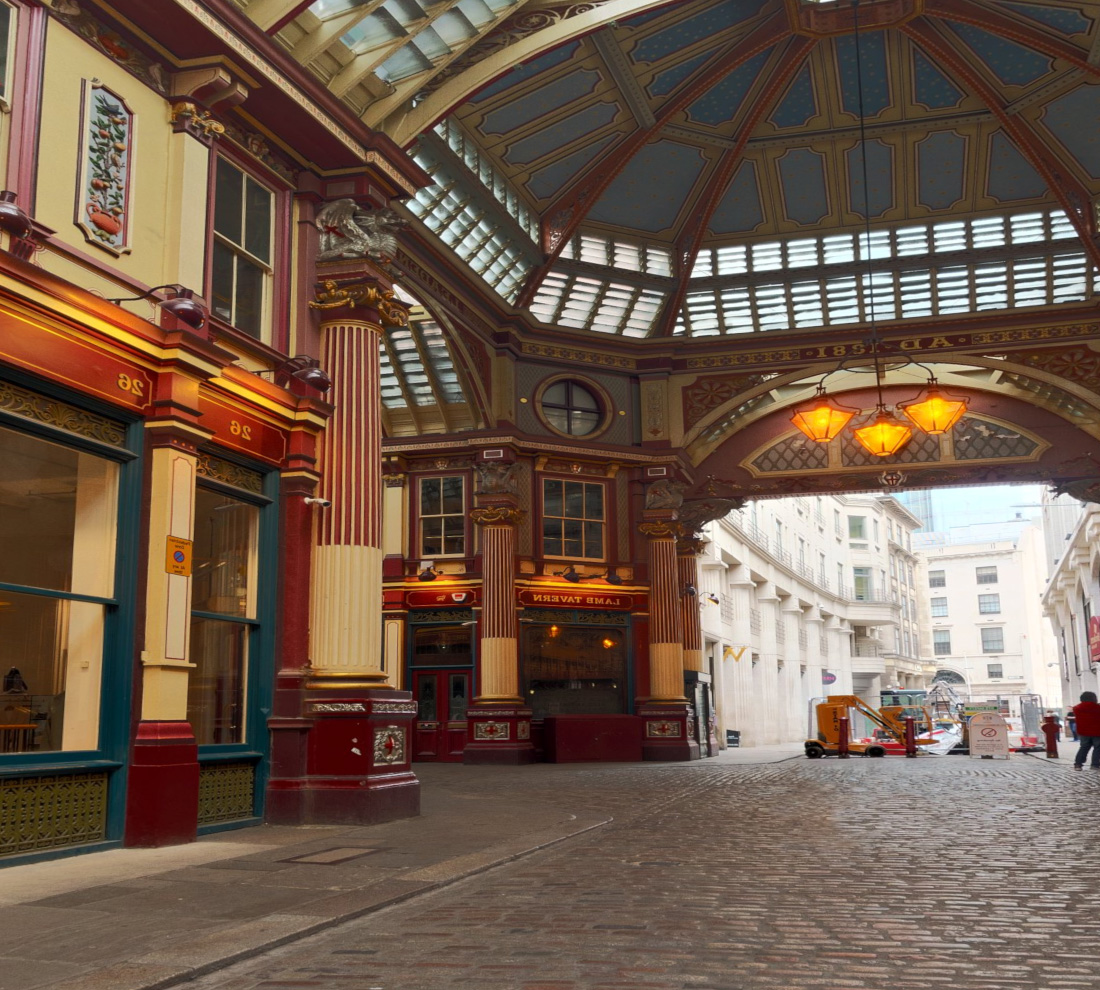}
\captionsetup{justification=centering}
\caption{First iteration.}
\label{fig_images/market_pass0}
    \end{subfigure}
    \hfill
    \begin{subfigure}[t]{0.46\linewidth}
\includegraphics[width=\linewidth]{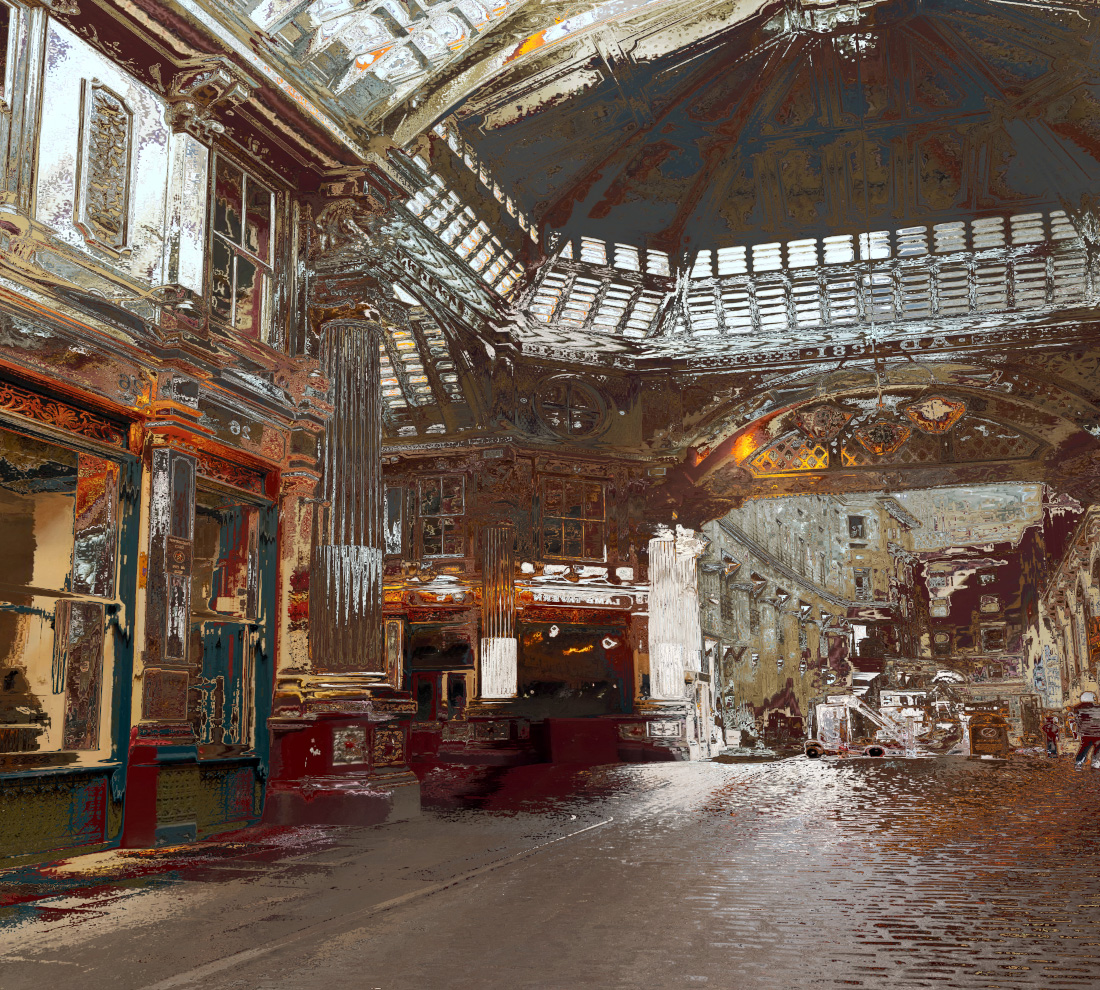}
\captionsetup{justification=centering}
\caption{Second iteration.}
\label{fig_images/market_pass1}
    \end{subfigure}
    \hfill
    \begin{subfigure}[t]{0.46\linewidth}
\includegraphics[width=\linewidth]{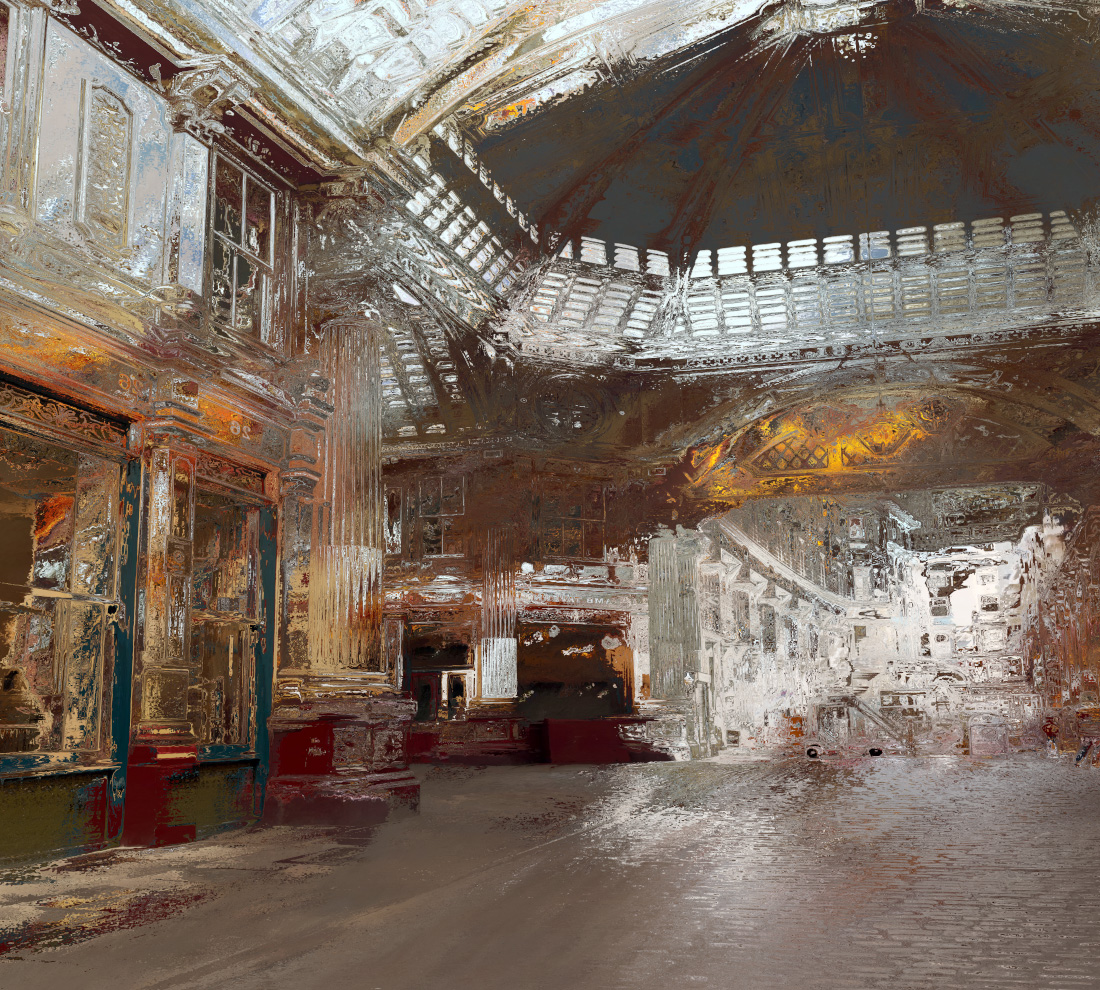}
\captionsetup{justification=centering}
\caption{Third iteration.}
\label{fig_images/market_pass2}
    \end{subfigure}
    \hfill
    \begin{subfigure}[t]{0.46\linewidth}
\includegraphics[width=\linewidth]{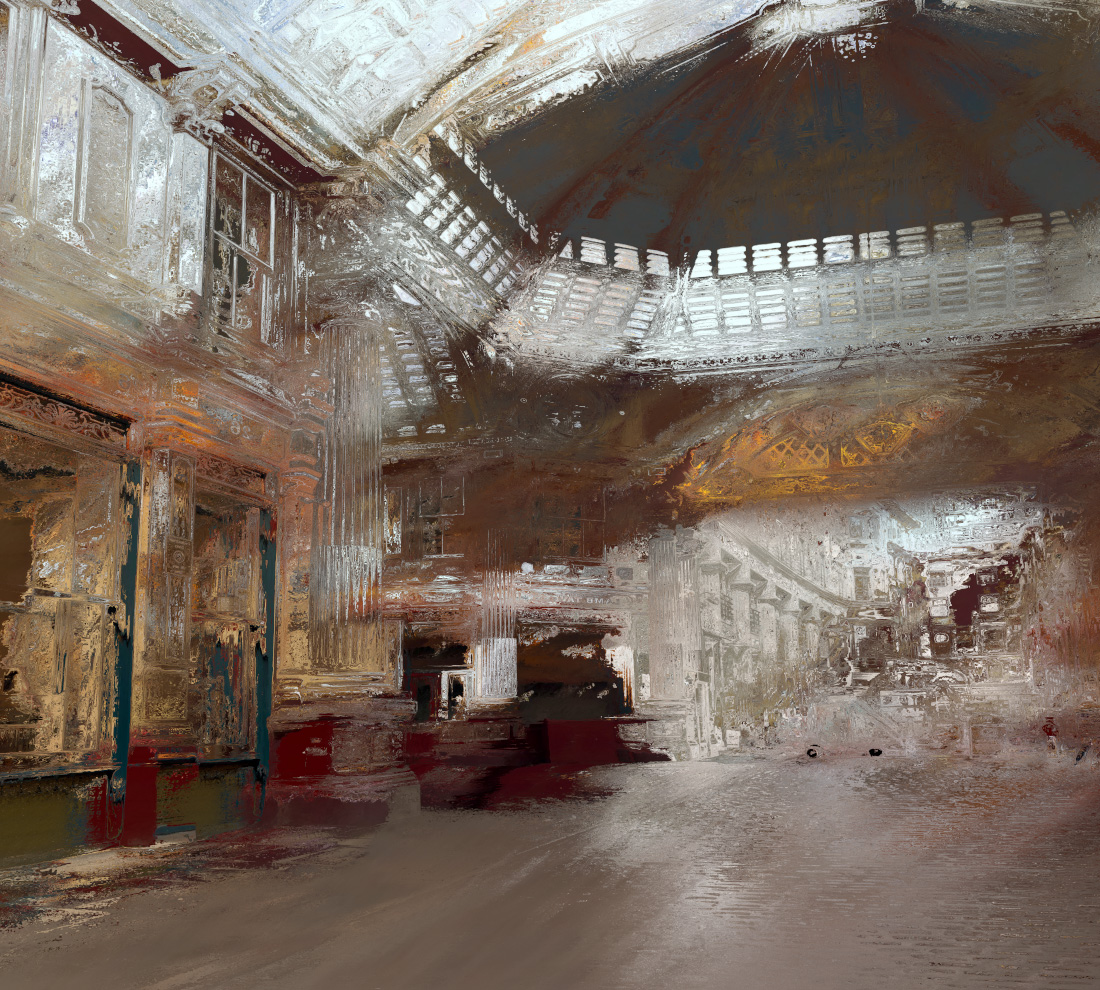}
\captionsetup{justification=centering}
\caption{Forth iteration.}
\label{fig_images/market_pass3}
    \end{subfigure}
    \hfill
    \begin{subfigure}[t]{0.46\linewidth}
\includegraphics[width=\linewidth]{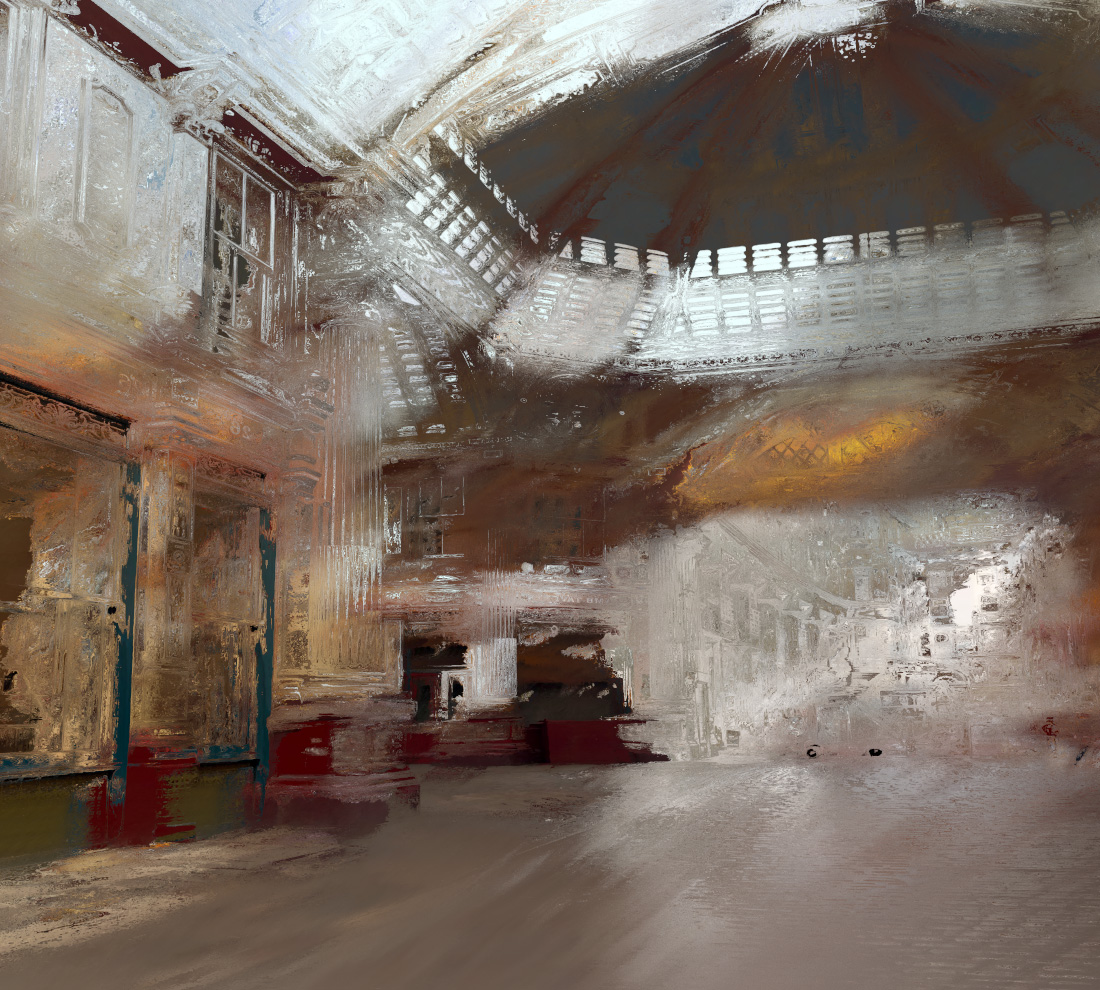}
\captionsetup{justification=centering}
\caption{Fifth iteration.}
\label{fig_images/market_pass4}
    \end{subfigure}
    \hfill
    \begin{subfigure}[t]{0.46\linewidth}
\includegraphics[width=\linewidth]{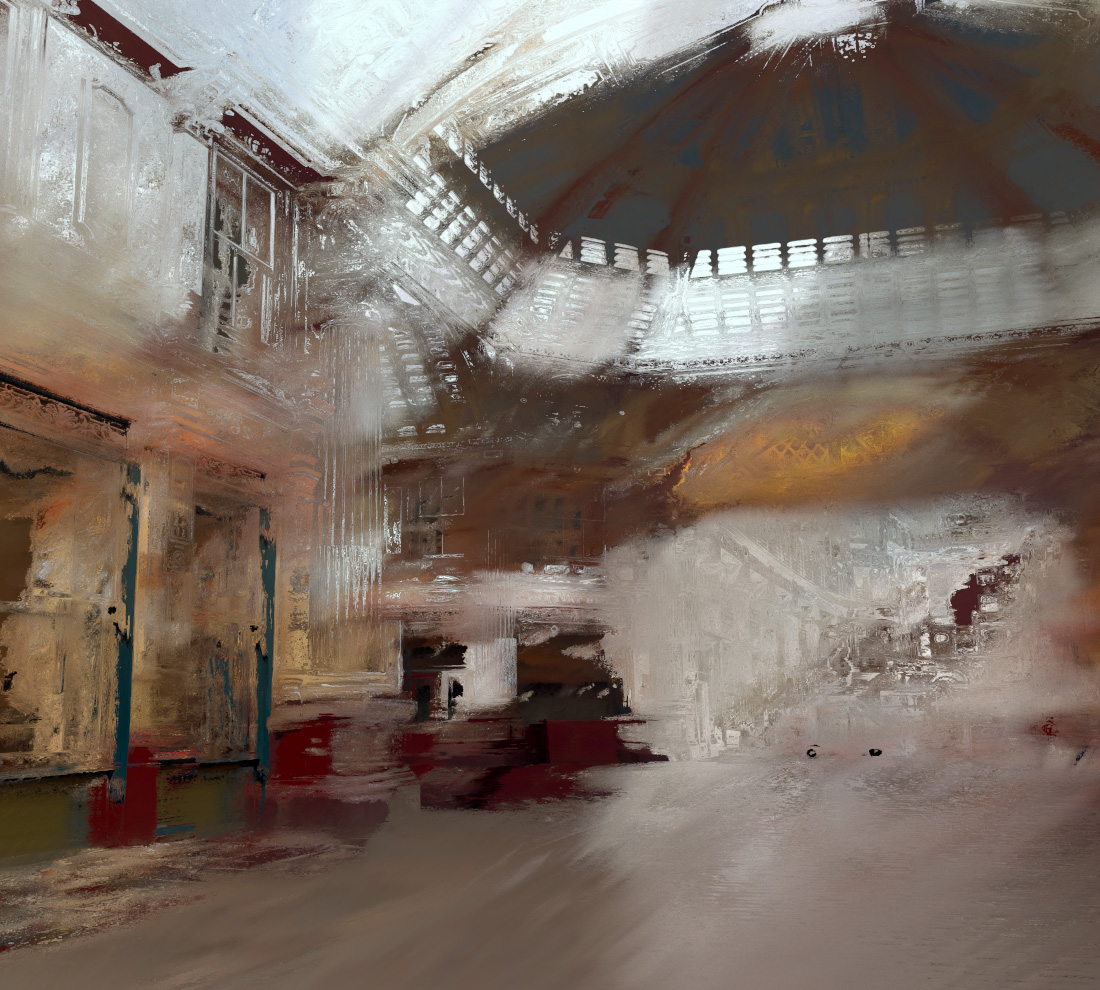}
\captionsetup{justification=centering}
\caption{Sixth iteration.}
\label{fig_images/market_pass5}
    \end{subfigure}
\caption{\it Recursion process that demonstrates more and more smudging. }
\label{fig_images/market_pass}
\end{figure*} 

\begin{figure*}[thbp!]
\includegraphics[width=0.49\linewidth]{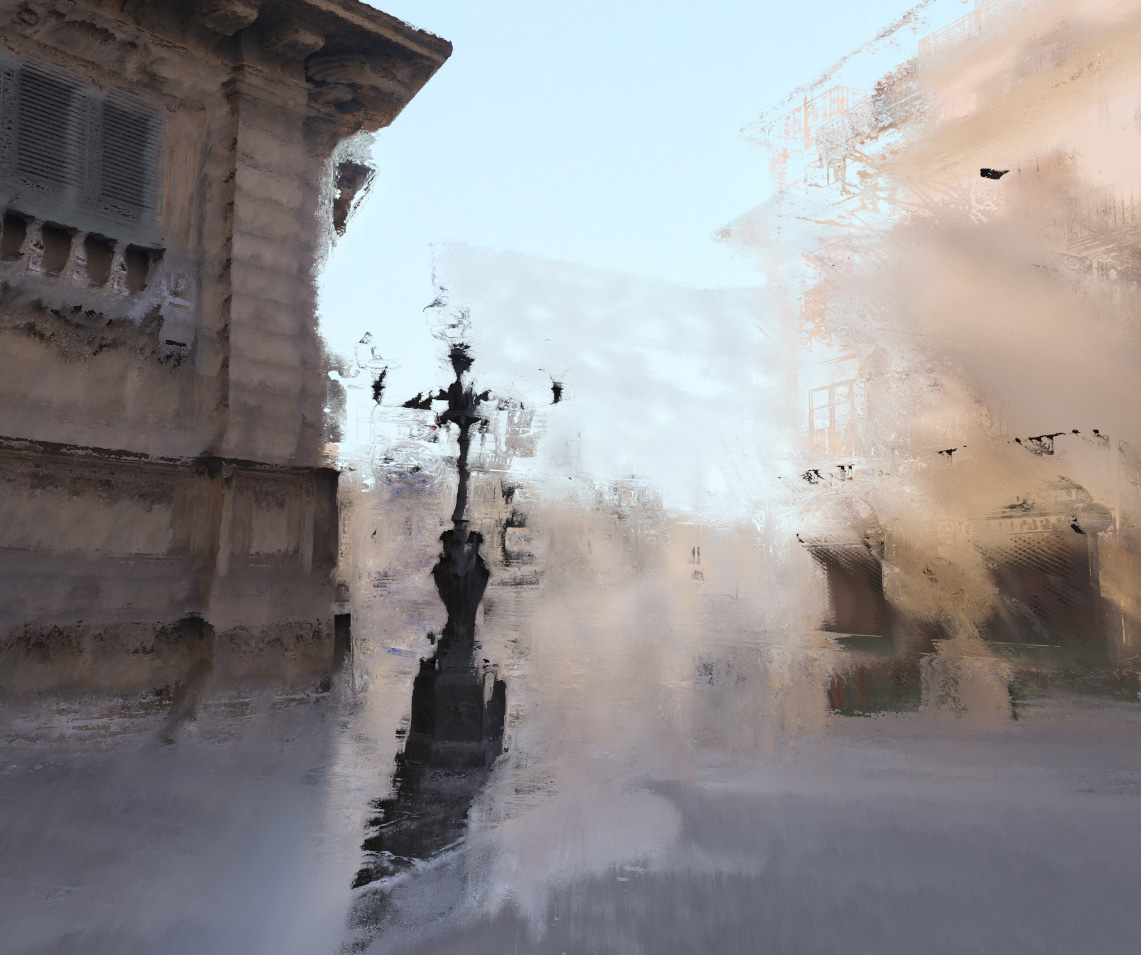}\hfill
\includegraphics[width=0.49\linewidth]{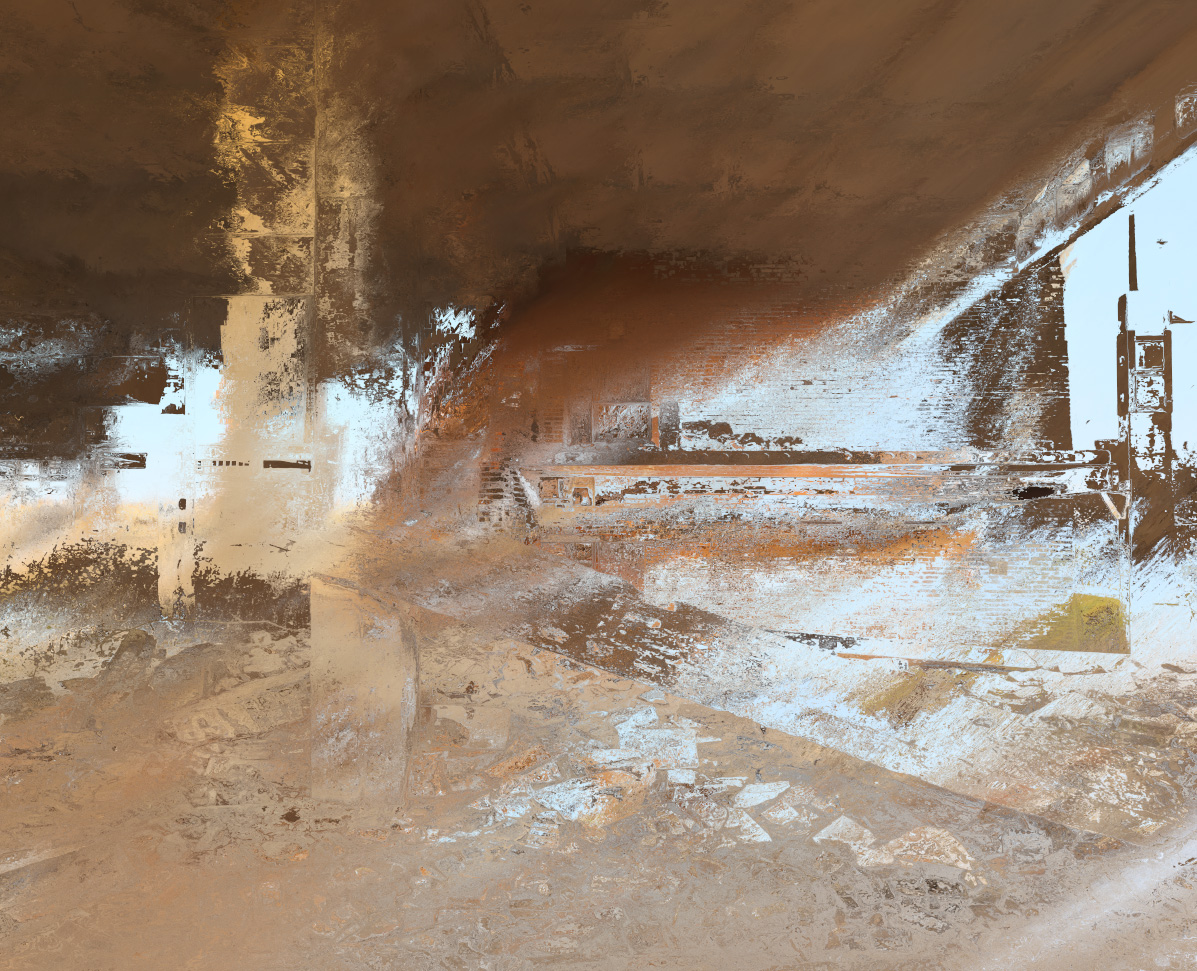}\hfill
\vspace{0.05in}
\includegraphics[width=0.49\linewidth]{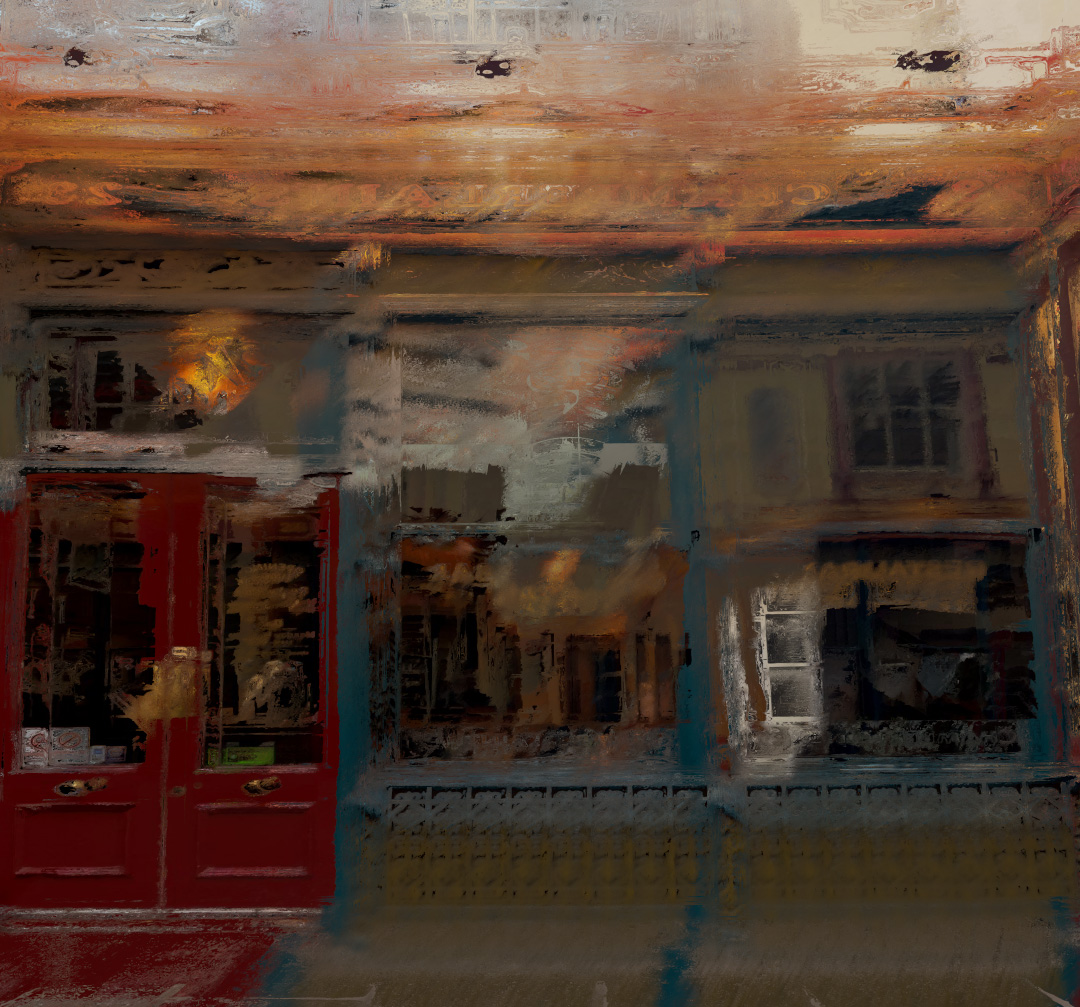}\hfill
\includegraphics[width=0.49\linewidth]{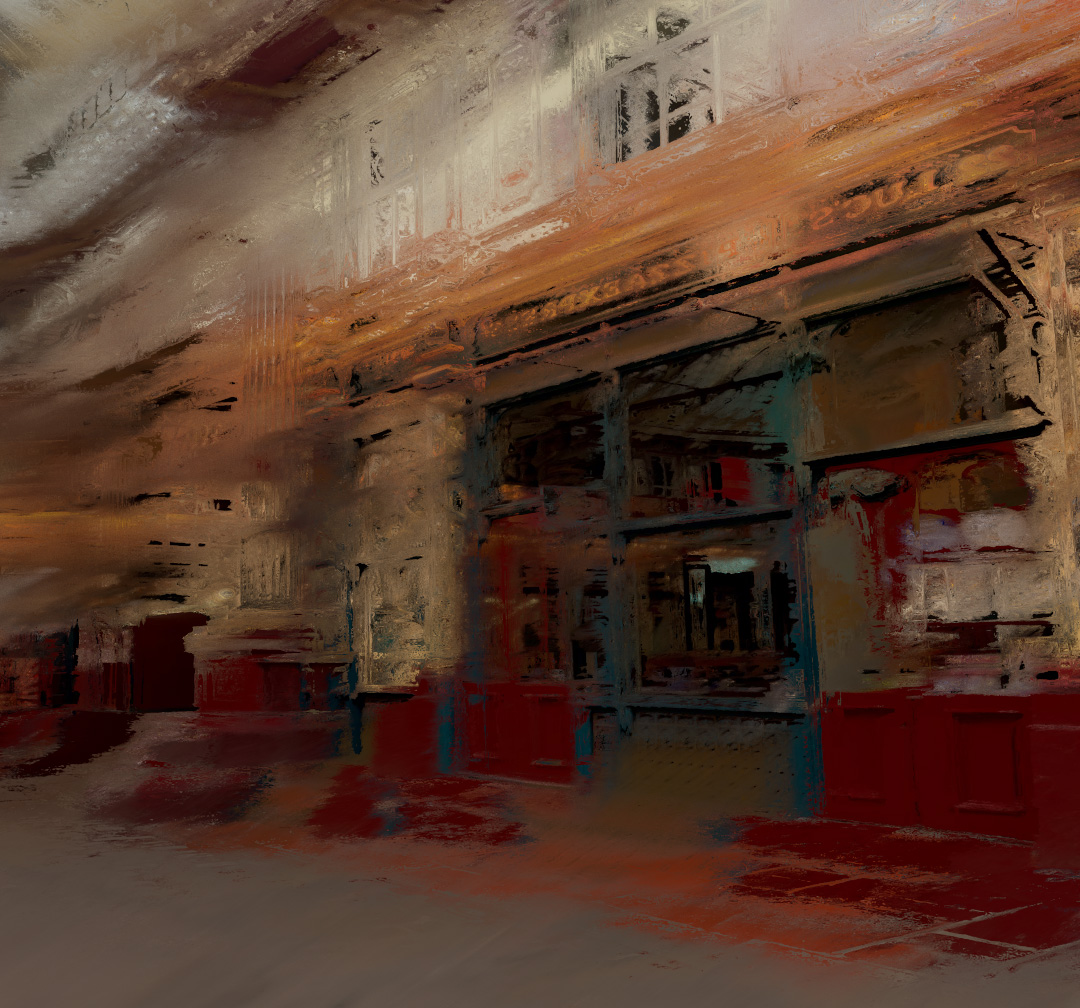}\hfill
\caption{\it More examples of GPU-based ray tracing with color-based displacements. These are six-iteration images that were captured during a real-time walk-through. }
\label{fig_raytracing}
\end{figure*}

\begin{figure*}[thbp!]
\includegraphics[width=0.49\linewidth]{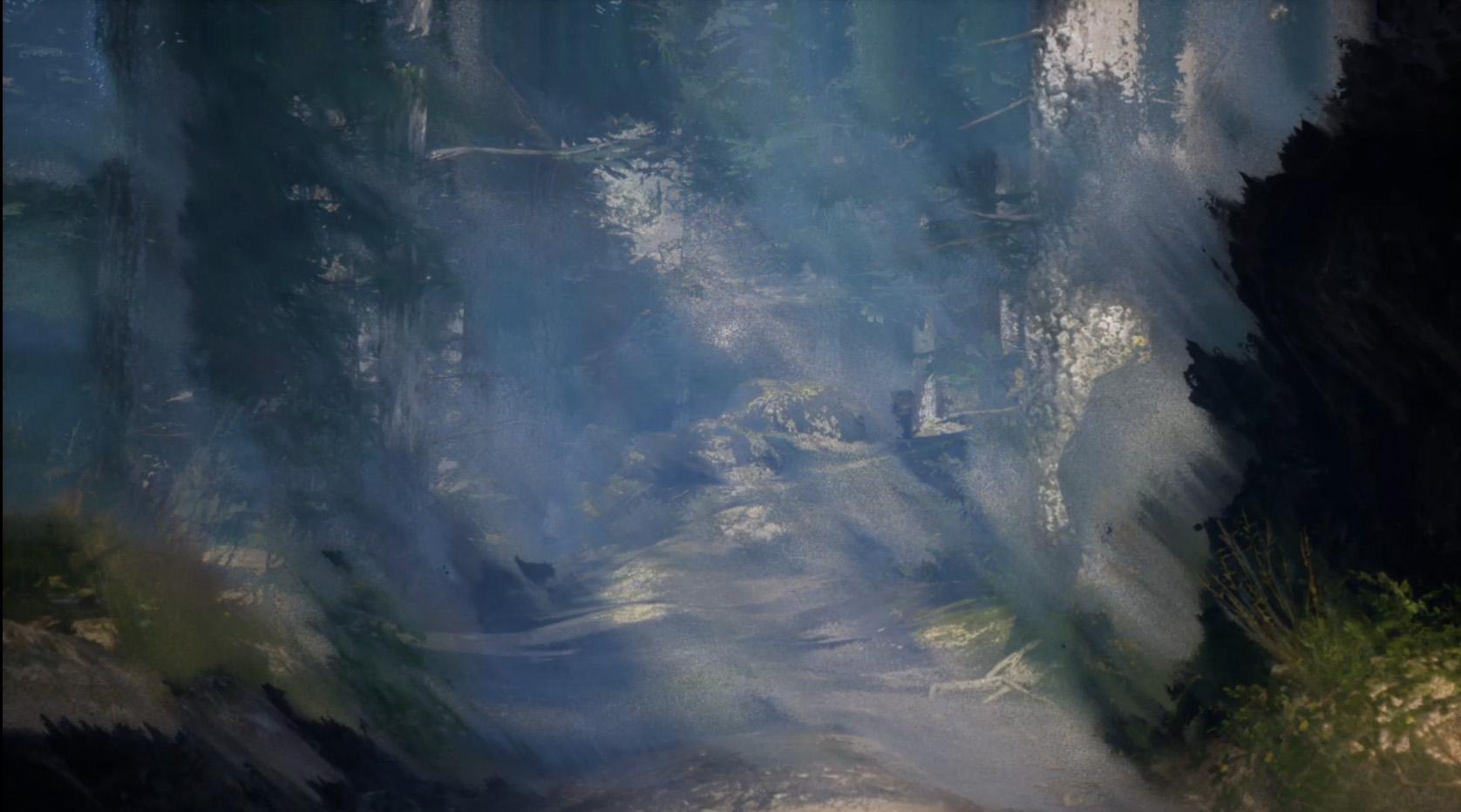}\hfill
\includegraphics[width=0.49\linewidth]{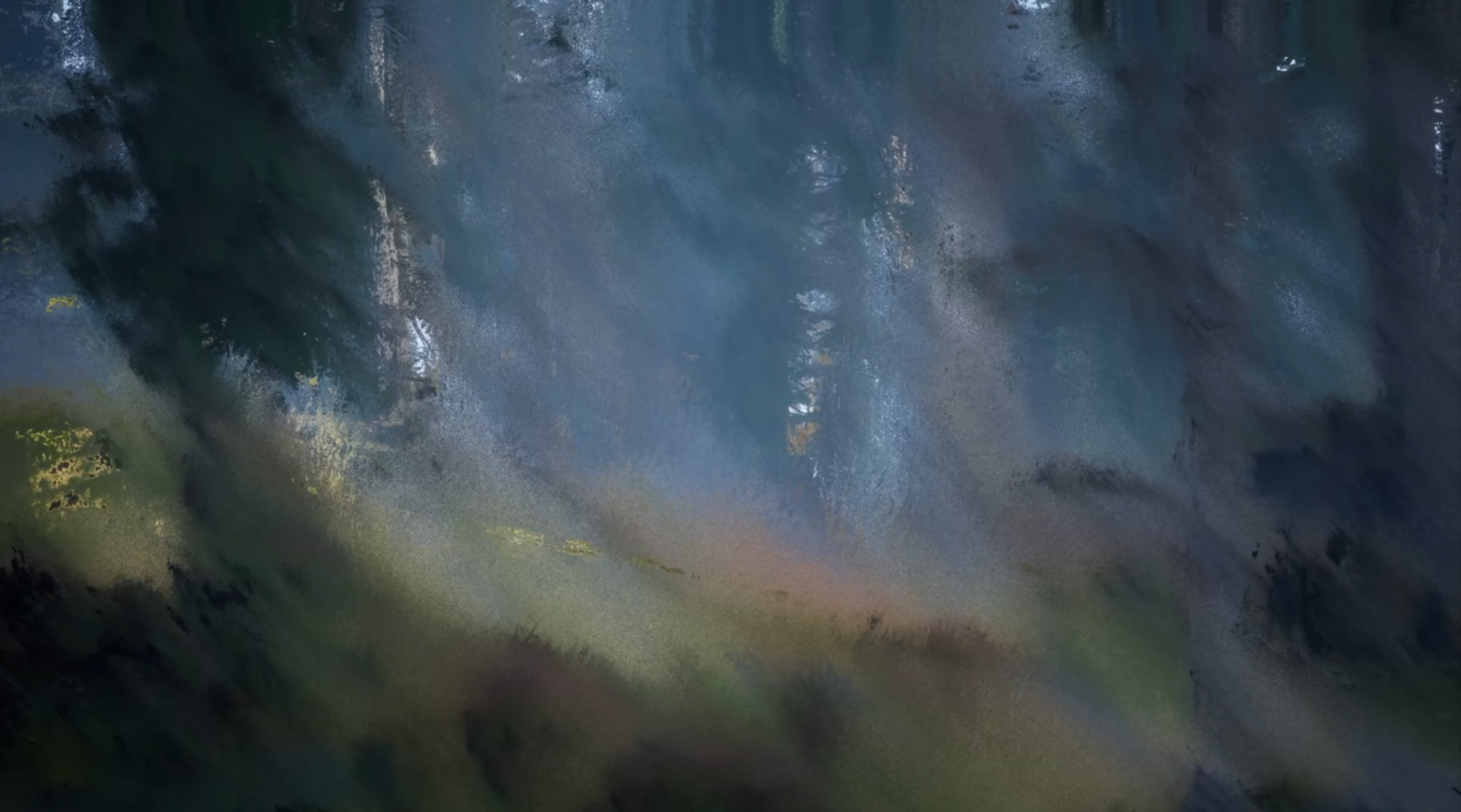}\hfill
\vspace{0.05in}
\includegraphics[width=1.00\linewidth]{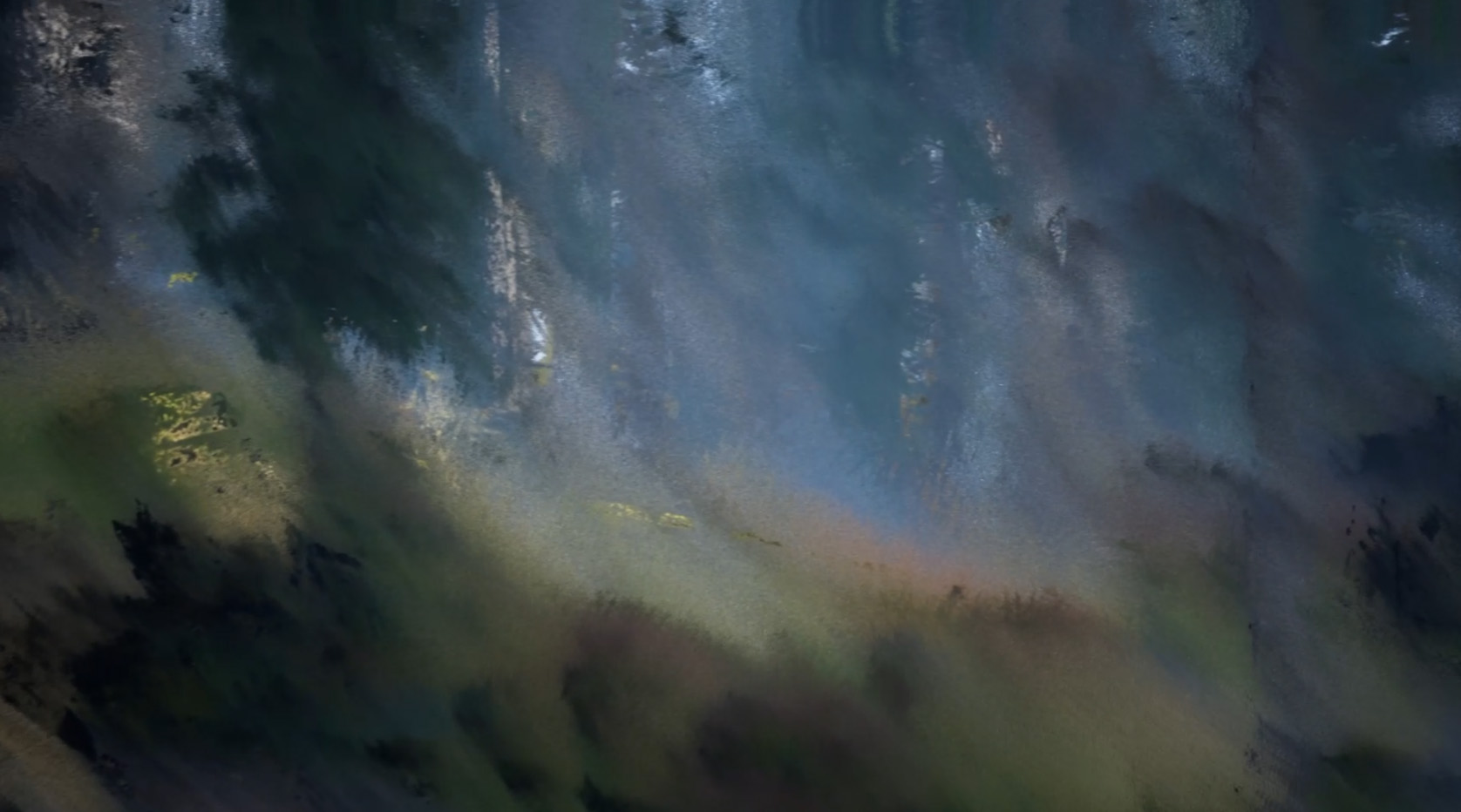}\hfill
\caption{\it Examples of unreal engine renderings using color-based displacements. These are also six-iteration images that are captured in a real-time walk-through. Note how much this forest resembles the forests in Disney's animations. }
\label{fig_unreal0}
\end{figure*} 

\begin{figure*}[thbp!]
\includegraphics[width=1.0\linewidth]{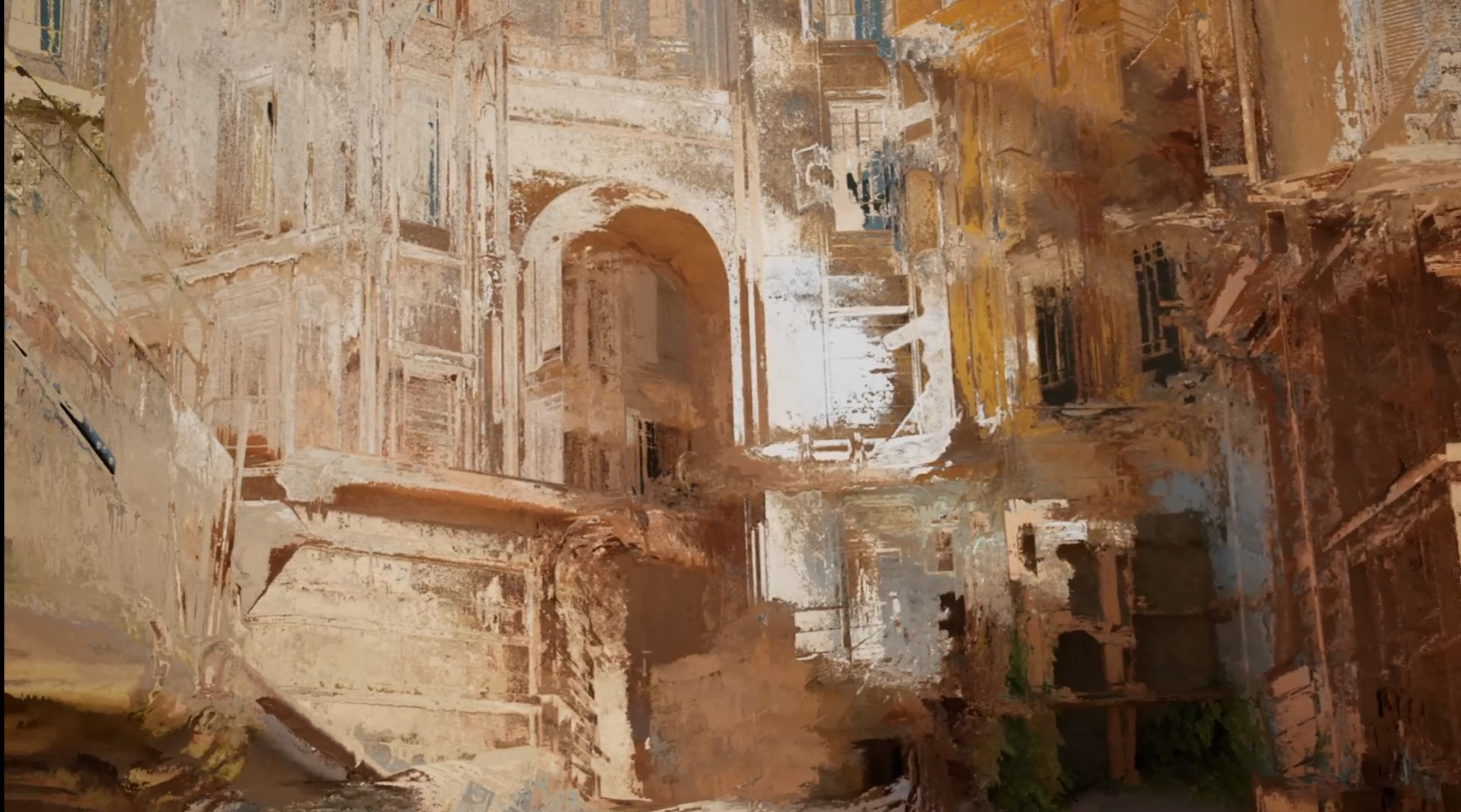}\hfill
\vspace{0.05in}
\includegraphics[width=0.497\linewidth]{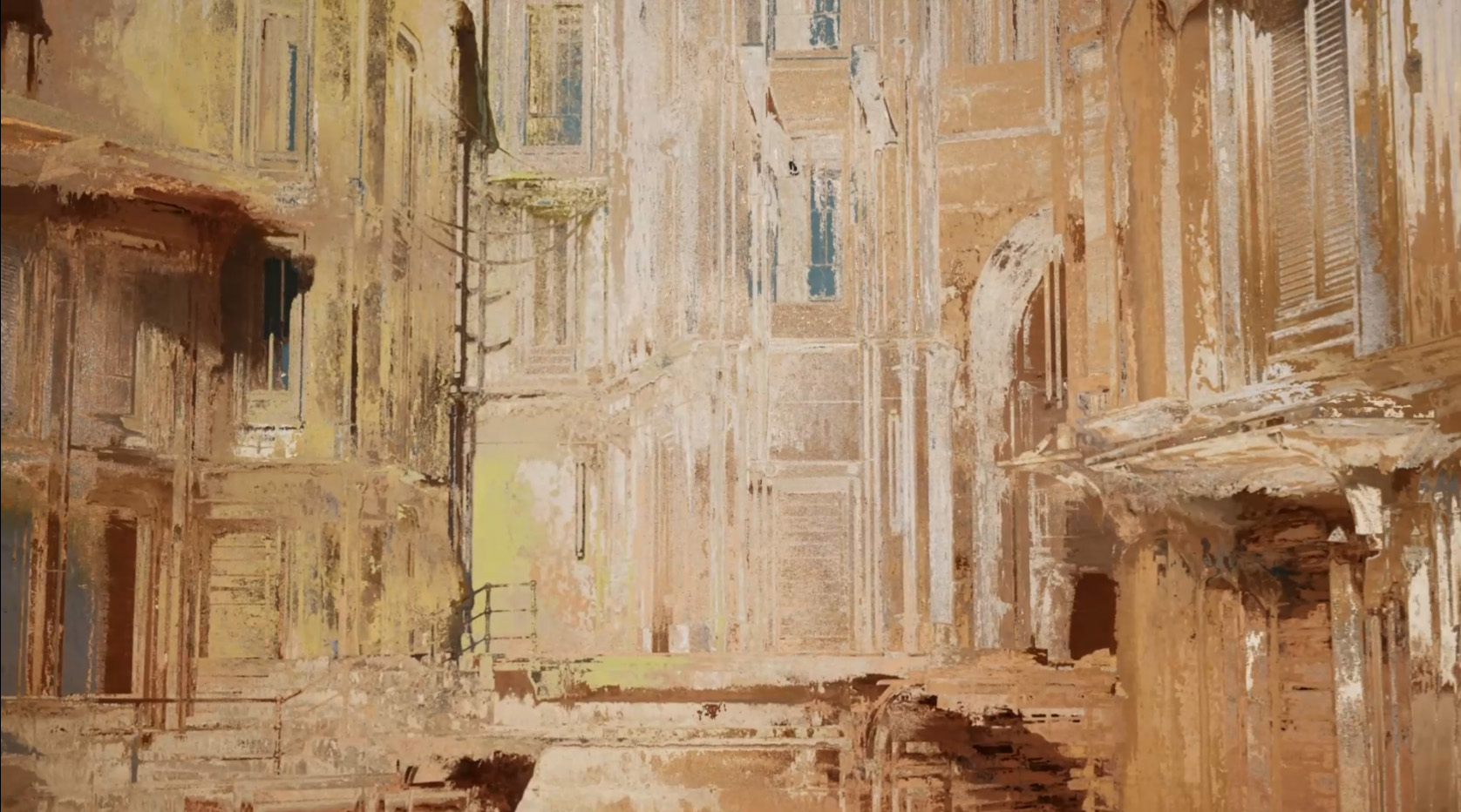}\hfill
\includegraphics[width=0.497\linewidth]{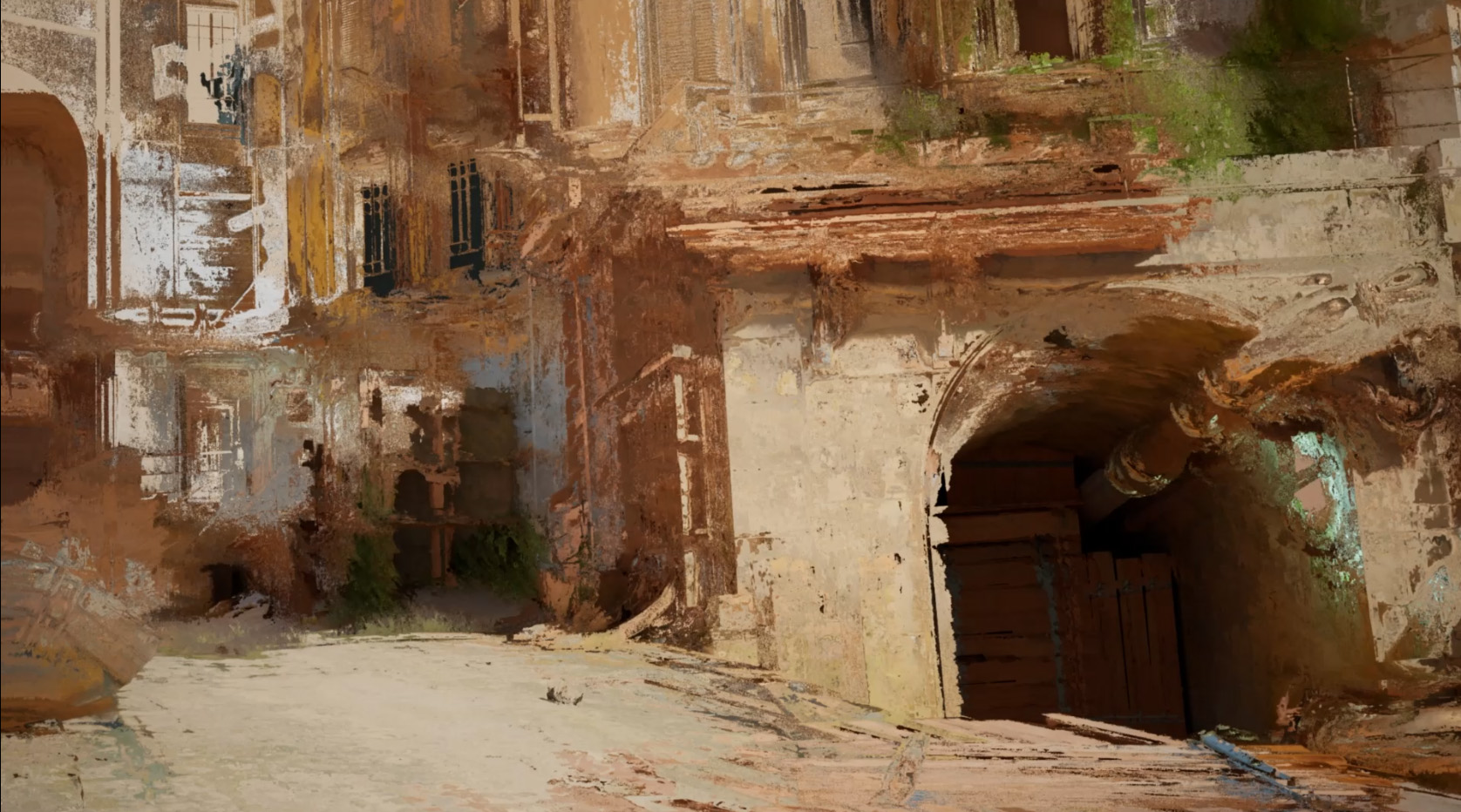}\hfill
\caption{\it Examples of unreal engine renderings using color based displacements. These are also six-iteration images that are captured in a real-time walk-through. }
\label{fig_unreal1}
\end{figure*}

Multi-view perspective is also an important tool in traditional animation to create moving cameras \cite{glassner2000}. There exists a significant number of techniques to obtain multi-view perspective such as creating cameras using parametric surfaces \cite{glassner2000,smith2004,yu2004framework}, using images to control camera parameters \cite{meadows2000b,morrison2020remote}, and using high-saliency regions to obtain cubist looking images \cite{collomosse2003cubist,arpa2012perceptual,wang2011cubist}. 

One of the problems of multi-view rendering is to avoid the shower door effect, which is a well-known problem in non-photo-realistic animations \cite{meier1996painterly}. Avoiding the shower-door effect requires dynamically controlling the parameters of all cameras during multi-view animations. Recently, another method of camera painting was developed, ``Remote Empathetic Viewpoint'', to avoid the effect of the shower door by decomposing the scene using control lights \cite{morrison2020remote}. However, this system is not real-time, and the control lights decompose the scene in a predictive way. Moreover, it is hard to control the style using ``Remote Empathetic Viewpoint''. 

In this paper, we present a new approach that can remove shower door effects by using rendered data to control camera parameters. An additional advantage of this approach is that the recursive application of this process creates painterly blurs that are obtained by deliberately mixing charcoal and oil paint to smudge the brighter colors that resemble the styles of a wide variety of representational and abstract artists \cite{hess2004willem}. 

\section{Theoretical Framework \label{fig_TF}} 

Let $I_n(u,v)$ denote the image/data rendered in time step $n$ and $(u,v)$ denote the texture coordinates of the image. Note that we are using $(u,v)$ instead of pixels. In this way, we can compute any number of samples per pixel using standard jittering to avoid aliasing. $I_n(u,v)$ can provide not only color information but also other data such as z-depth, shading normal, surface normal, material properties of the shading point, or object ID for the given coordinate. Our recursive process is straightforward and can be represented as $$I_n(u,v)= F(I_{n-1}(u,v), S(n)),$$
where $I_{n-1}(u,v)$ denote the image / data rendered previously and $S(n)$ is a given scene in the time step $n$.

$F$ is a function that represents a camera-painting process that defines a new camera parameter per pixel. The process $F$ consists of two stages: 
\begin{enumerate}
\item Assigning a camera to each point $u,v$ based on previously computed image/data $I_{n-1}(u,v)$ and the current scene description $S(n)$; 
\item For every $u,v$ compute the new image/data using new camera definition and $S(n)$.
\end{enumerate}
The camera assignment, in general, is simply defining a ray that is usually provided by an eye point that is given by its position $\mathbf{P_e}(u,v)$ and a direction that is given a unit vector $\vec{N}(u,v)$. For simplicity, the unit vector is computed indirectly from a second position $\mathbf{P_p}(u,v)$ that corresponds to the 3D position of the texture coordinate $(u,v)$. Then,  
$$\vec{N}(u,v) = \frac{\mathbf{P_p}(u,v)- \mathbf{P_e}(u,v)}{|\mathbf{P_p}(u,v)- \mathbf{P_e}(u,v)|}$$
We compute the new 3D positions of the eye(s) and 3D positions of the texture coordinates from their original positions using two displacement vectors, $\vec{V_e}(u,v)$ and $\vec{V_p}(u,v)$ as $$\mathbf{P'_p}(u,v)=\mathbf{P_p}(u,v) + \vec{V_e}(u,v)$$
$$\mathbf{P'_p}(u,v)=\mathbf{P_p}(u,v) + \vec{V_p}(u,v)$$
Note that the whole problem reduces the development of mappings to compute the displacement vectors $\vec{V_e}(u,v)$ and $\vec{V_p}(u,v)$ from $I_{n-1}(u,v)$. This provides significant flexibility for artists to construct their mappings. In the next section, we will provide a specific example that uses only colors in $I_{n-1}(u,v)$.

\section{Displacement Using Colors}

Now, let us demonstrate the process using a simple mapping that only involves color to define cameras and uses a constant $\mathbf{P_e}$, i.e. $\vec{V_e}(u,v)=(0,0,0)$. Now, only the new positions of 3D texture coordinates, $\mathbf{P'_p}(u,v)$, need to be produced to create a new camera. Let $$I_{n-1}(u,v) = (R_{n-1}(u,v), G_{n-1}(u,v), B_{n-1}(u,v))$$ denote the three channels of low dynamic range that are previously computed in time $n-1$. Let us assume that the vector $\vec{V_p}(u,v)$ is computed as follows:
$$
\vec{V_p}(u,v) = \lambda \left( 
 \begin{matrix}
2 R_{n-1}(u,v) -1\\
2 G_{n-1}(u,v) -1 \\
2 B_{n-1}(u,v) -1\\
\end{matrix}
\right)
$$
where $\lambda$ is a positive real number and $2 I_{n-1}(u,v) -1) \in (-1,1)^3$. Note that $\lambda (2 I_{n-1}(u,v) -1))$ is a vector in $(-\lambda,\lambda)^3$ in the local coordinate system of the camera.

This process can provide a new camera for any given $(u,v)$. We can then compute the color of $I_n(u,v)$ by shooting rays from $\mathbf{P_e}$ to $\mathbf{P'_p}(u,v)$ for a given scene at time $n$, $S(n)$. We apply the process $n$ times starting from $I_{0}(u,v) = (0.5,0.5,0.5)$. Note that this gives us a traditional ray-traced image in the first iteration (see Figure~\ref{fig_images/market_pass0}). The image computed in $I_{1}(u,v)$ determines how the next image is computed. For example, $R=1$ moves the pixel position to the right, and $R=0$ to the left in the local camera coordinate system. Similarly, $G=1$ moves the position of the pixels to the upward, $G=0$ to the downward in the local camera coordinate system. Blue does not have too much of an effect. 

An important property of this process is that average color values do not move the rays. Only extreme values move the rays. Moreover, in regions closer to a boundary, the ray moves up and down, or right and left, based on the values in each iteration. This process produces smudges by mixing extreme colors and produces an effect that is similar to painterly smudges. 

This particular smudging effect is different from any type of anisotropic blur that can be obtained by bilateral filters or line integral convolution since the process is not entirely predictable: it allows sharp regions while blurring other regions. The iterative process can be shown in \ref{fig_images/market_pass}.

Note that even using colors to create displacements is general enough. We can choose to display any particular color by simply using more complicated mappings in the form of
$$
\vec{V_p}(u,v) = W \left(R_{n-1}(u,v),G_{n-1}(u,v),B_{n-1}(u,v)
\right)
$$

\begin{figure*}
\includegraphics[width=0.49\linewidth]{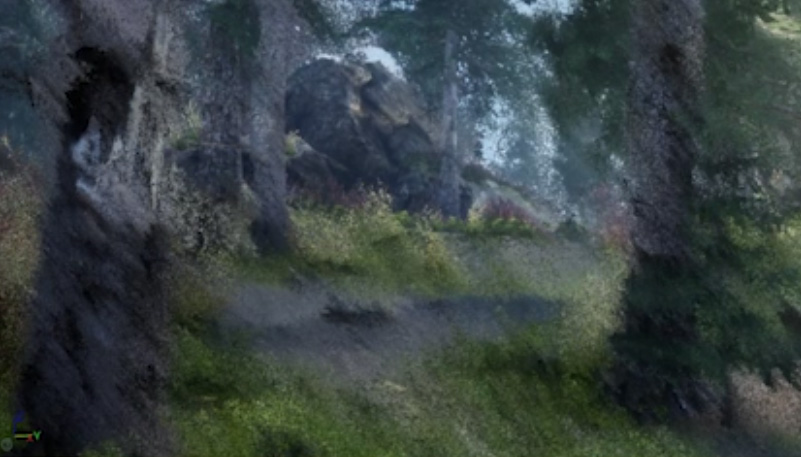}\hfill
\includegraphics[width=0.49\linewidth]{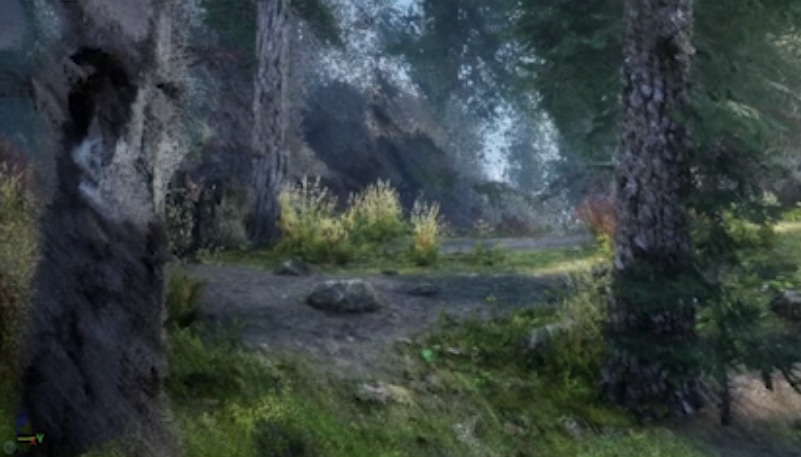}\hfill
\vspace{0.05in}
\includegraphics[width=1.00\linewidth]{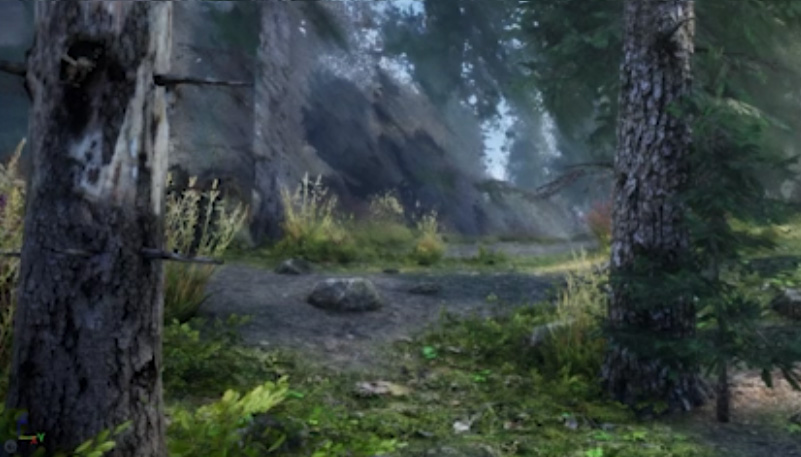}\hfill
\caption{\it Changing painterly focus by controlling displacements with depth. These are also six-iteration images created using Unreal. }
\label{fig_unrealfocus}
\end{figure*}

\begin{figure*}
\includegraphics[width=0.49\linewidth]{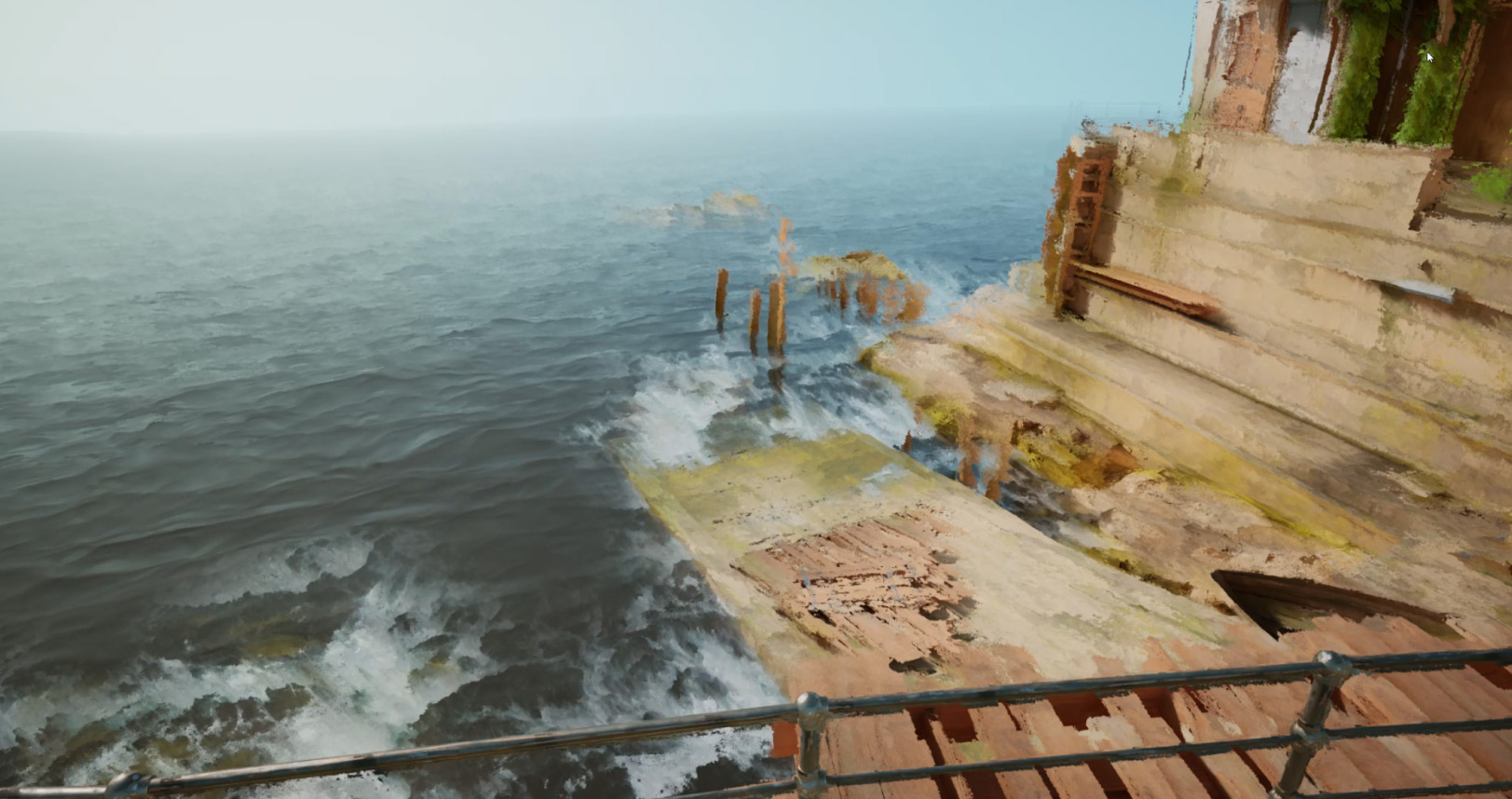}\hfill
\includegraphics[width=0.49\linewidth]{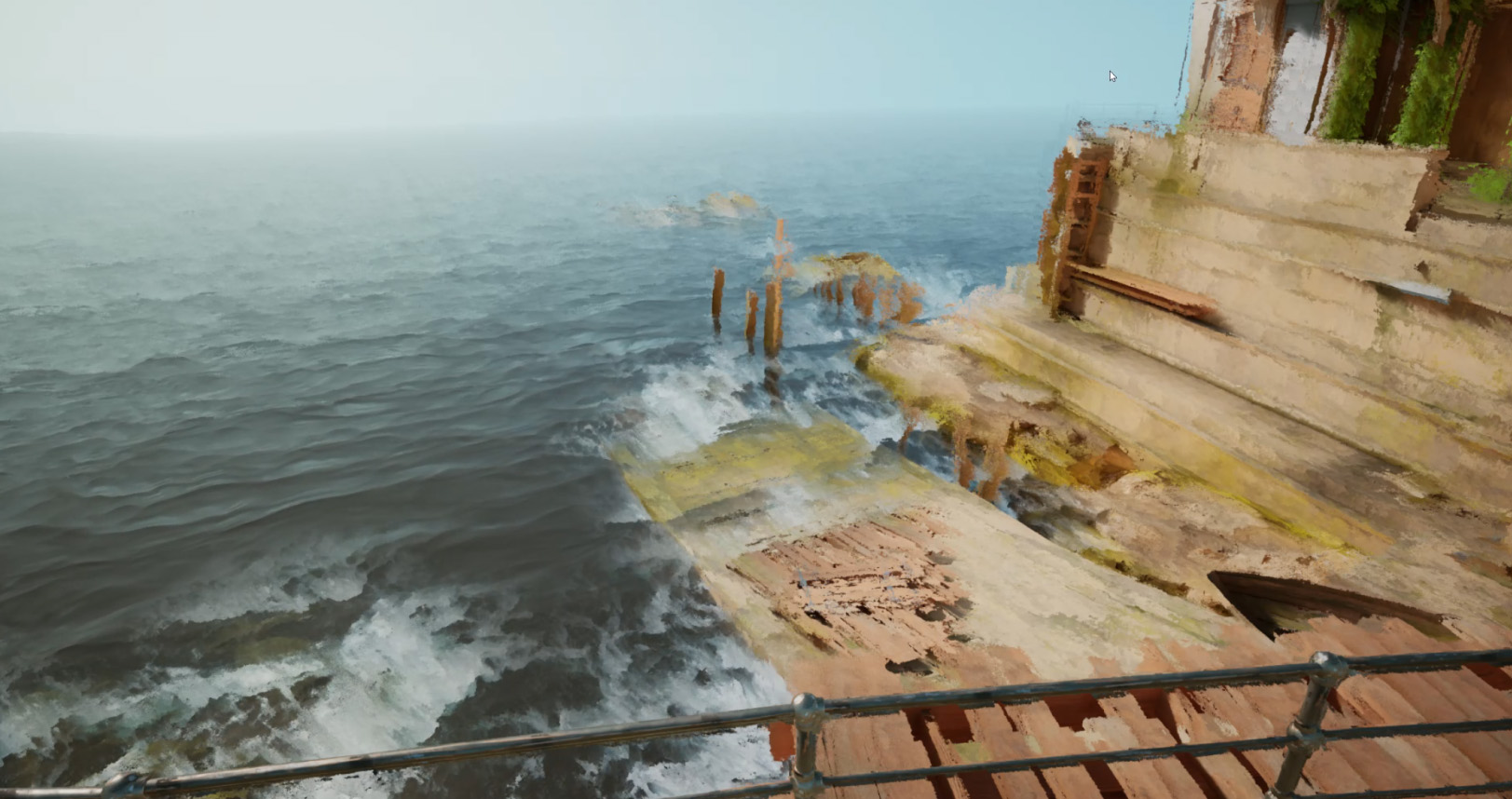}\hfill
\vspace{0.05in}
\includegraphics[width=1.00\linewidth]{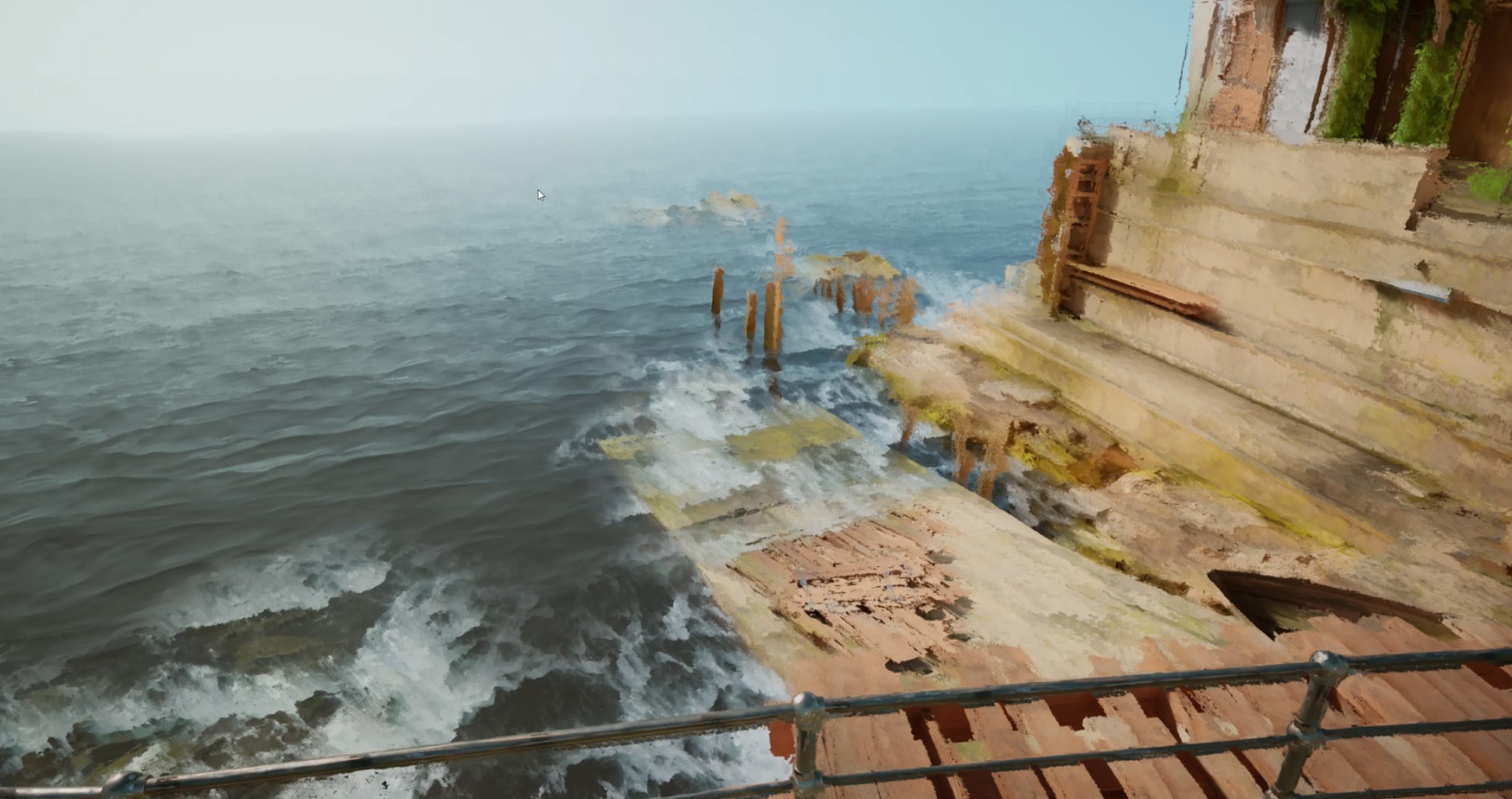}\hfill
\caption{\it A painterly focus scene with moving sea waves. These are also six-iteration images created using Unreal. }
\label{fig_unrealwaves}
\end{figure*}

\section{Displacements Based on Depth}

As discussed in Section~\ref{fig_TF}, the z-depth can also be used to control displacements. Let us say that we want to create an illusion similar to out of focus by controlling painterly smudge based on depth. Let $z(u,v)$ denote the depth information for a given texture coordinate, and let $z_0$ denote the depth that needs to be in focus. We can use $|z - z_0|$ as a multiplier as follows:
$$
\vec{V_p}(u,v) = \lambda |z - z_0| \left( 
 \begin{matrix}
2 R_{n-1}(u,v) -1\\
2 G_{n-1}(u,v) -1 \\
2 B_{n-1}(u,v) -1\\
\end{matrix}
\right)
$$
Using this approach, we obtain visuals that resemble 2D animation aesthetics in 3D. This is because in 2D animations the backgrounds are drawn in a painterly way and characters are drawn sharp. 

\section{Displacement Based on Normals}
Another way to control displacements is to use world normals and camera direction. In this case we would replace the usual color values, $R_{n-1}(u,v)$, $G_{n-1}(u,v)$, and $B_{n-1}(u,v)$, with the cross product of the camera direction $\vec{C}(u,v)$ and the world normals $\vec{N}(u,v)$.
$$
\vec{V_p}(u,v) = \lambda (\vec{C}(u,v) \times \vec{N}(u,v))
$$

The normal of a surface is a vector perpendicular to the surface itself, pointing outward from the shape. So taking its cross product with camera direction results in a tangent or a vector parallel to the surface. Small adjustments to this equation can have many widely varied effects. For example, looping through the equation created images that looked layered. Adding two different kinds of randomness, on the other hand, resulted in one version that looks fractured and another that looks like cross-hatching. 

\section{Implementations} 

We have first implemented this approach using the Nvidia OptiX Engine \cite{parker2010optix}. In the examples in Figures~\ref{fig_teaser}, \ref{fig_images/market_pass}, and \ref{fig_raytracing}, we used simple scenes that consist of only a spherical texture mapped onto an infinite sphere. The process also handles 3D scenes in real-time. We have also created examples with complex 3D objects in videos captured in real time. 

One problem with the ray tracing solution is finding high-quality content, i.e. very richly designed scenes. Therefore, we turned our attention to game engines. Now, there exists a significant amount of Virtual Reality content developed for Game Engines such as Unreal \cite{sanders2016introduction}. The problem with game engines is that they do not support real-time ray tracing. We then observed that the $\vec{V_e}(u,v)=(0,0,0)$ solution does not require Ray Tracing. If we keep $\mathbf{P_e}$ in the same position, the rendered image already includes most of the potential directions. Then, our original mapping reduces into an image processing problem. We can simply pick colors from another position in the image on the basis of the displacement. This process can be done in the Unreal Engine in real-time. We can also access other information such as z-depth in the Unreal Engine. 

In conclusion, we have implemented the color- and depth-based displacement methods in Unreal Engine. We tested with a wide variety of 3D scenes that are created by Unreal users with real-time walk-throughs.  Figures~\ref{fig_unreal0}, ~\ref{fig_unreal1}, \ref{fig_unrealfocus}, and \ref{fig_unrealwaves} show frames from these walks-through.

 \begin{figure*}[thbp!]
    \begin{subfigure}[t]{0.48\linewidth}
\includegraphics[width=\linewidth]{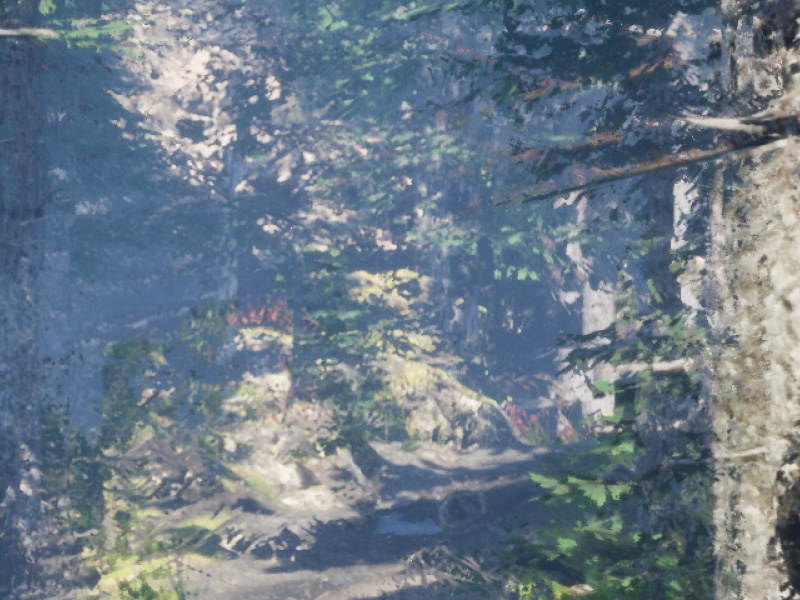}
\captionsetup{justification=centering}
\caption{Version 1 - Smudge Direction calculated using World Normals.}
\label{fig_images/Forest009.jpg}
    \end{subfigure}
    \hfill
    \begin{subfigure}[t]{0.48\linewidth}
\includegraphics[width=\linewidth]{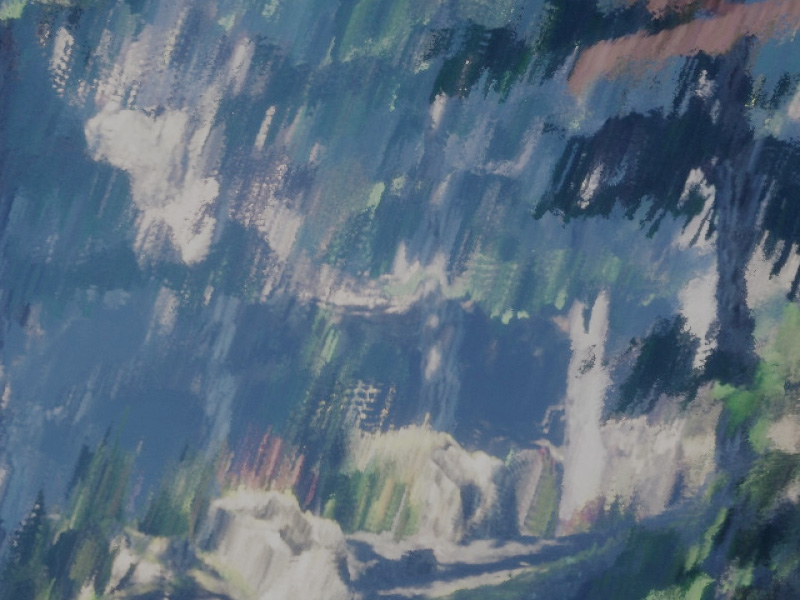}
\captionsetup{justification=centering}
\caption{Version 2 - Smudging from Version 1 is looped to run multiple times to create a layered effect.}
\label{fig_images/Forest010.jpg}
    \end{subfigure}
       \hfill
    \begin{subfigure}[t]{0.48\linewidth}
\includegraphics[width=\linewidth]{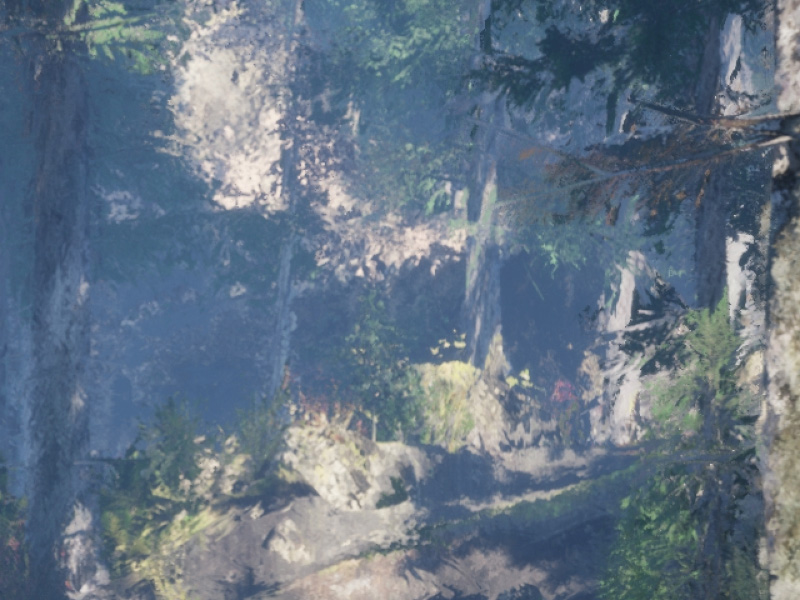}
\captionsetup{justification=centering}
\caption{Version 3 - A small amount of randomness is added using Unreal's Material Expression Vector Noise.}
\label{fig_images/Forest011.jpg}
    \end{subfigure}
       \hfill
    \begin{subfigure}[t]{0.48\linewidth}
\includegraphics[width=\linewidth]{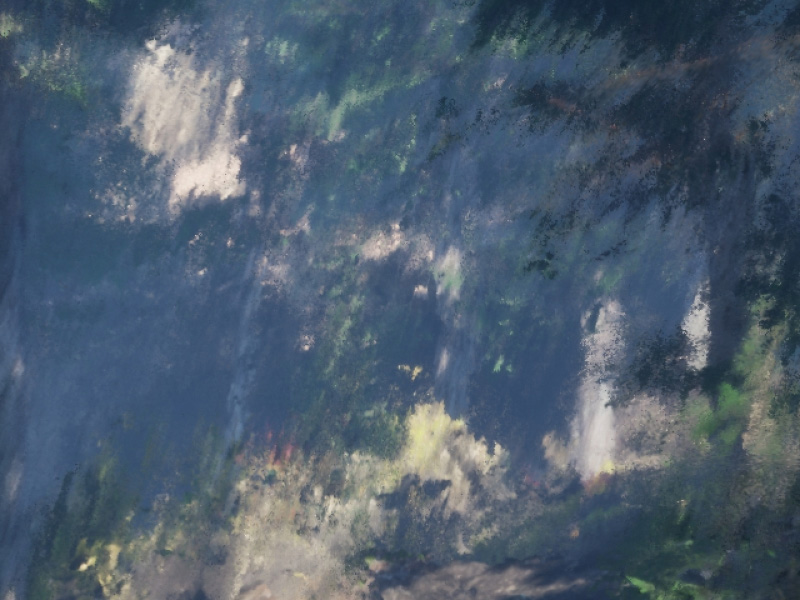}
\captionsetup{justification=centering}
\caption{Version 4 - Smudging using Normals and Noise from Version 3 is looped to run multiple times.}
\label{fig_images/Forest012.jpg}
    \end{subfigure}
\caption{\it Examples using world normals and camera direction to determine smudge direction.}
\label{fig_images/Forest2}
\end{figure*}

\subsection{Handling Edge of Screen Flickering}

One issue we originally had with our Unreal implementation was flickering at the edge of the screen. Because the smudging works by taking the neighboring pixel colors, when the original pixel is close to the edge of the screen, often the values being used to get the neighboring pixel were outside of the screen range. This resulted in unwanted glitching where Unreal attempted to compensate for the bad input. 

To fix this issue, we experimented with methods used in convolution filters that deal with the same problem. The first pass adjusted how the new uv coordinates are calculated. Now it uses the red and green channels of the current pixel as uv, multiplies by the maximum smudge distance, divides by the size of the screen, and finally adds the original uv to find the new pixel location. In the second pass, the edge values were clamped so that if the new uv position is less than 0 it becomes 0, and if it is greater than the screen size, it is set to the maximum screen size. This fixed the glitching but left unseemly bars of color at the edge of the screen that are very distracting. The next version wraps around instead of just clamping. In some cases, this produced a good result, but often when the top and bottom or left and right sides of the screen have very different colors, wrapping produces a very obvious border of color. 

Finally, we tried mirroring the values so that the new pixel color would be still somewhat near its original. The mirroring technique completely fixed the glitching seen at the edge of the screen and also produced the smoothest, least distracting edge result, as shown in Figure~\ref{fig_images/Forest2}. 

\section{Discussion and Future Work} 

In this work, we have demonstrated that recursive camera painting can provide a painting effect. We have scratched the surface in terms of controlling the results. We think that it will be possible to obtain a wider variety of effects such as cubist rendering by controlling cameras with the data collected from scenes. 
In terms of control, it is much better to use a real-time ray-tracer; however, the major bottleneck is finding high-quality scenes available. Currently, we only work with realistic materials and scenes. The results can be further improved by using nonphotorealistic shaders \cite{akleman2016} and nonrealistic scenes \cite{Elber2011}. 
An implication of implementation with Unreal Engine is that we can use a similar approach to manipulate real images during Augmented Reality. In other words, we can turn real scenes into painted environments in real-time using the same type of image processing operations.

\bibliographystyle{unsrtnat}
\bibliography{references,tamuthesis}

\end{document}